\documentclass[lettersize,journal]{IEEEtran}
\usepackage{amsthm}
\usepackage{cite}
\usepackage{amsmath,amssymb,amsfonts}
\usepackage{algorithm,algorithmic}
\usepackage{graphicx}
\usepackage{textcomp}

\usepackage{gensymb}
\usepackage{hyperref}
\usepackage{todonotes}
\usepackage{multirow}
\usepackage{booktabs}
\usepackage[nolist]{acronym} 
\usepackage{enumitem} 
\usepackage{tabularx,array}
\newcolumntype{Y}{>{\raggedright\arraybackslash}X}
\newcolumntype{M}{>{\centering\arraybackslash}X}
\newcolumntype{N}[1]{>{\raggedright\arraybackslash}p{#1}}
\newcolumntype{Z}{>{\centering\arraybackslash}p{10mm}} % narrow centered column
\newcolumntype{C}[1]{>{\centering\arraybackslash}p{#1}}

\usepackage{pgfplots}
\pgfplotsset{compat=1.9}
\usepackage{makecell}
\usepackage{threeparttable}

\newtheorem{theorem}{Theorem}
\newtheorem{lemma}[theorem]{Lemma}

\newtheorem{remark}{Remark}

\DeclareMathOperator{\R}{\mathbb{R}}

\DeclareMathOperator*{\argmin}{arg\,min}
\DeclareMathOperator{\N}{\mathbb{N}}

\newcommand{\change}[1]{\textcolor{black}{#1}}
\newcommand{\changeC}[1]{\textcolor{black}{#1}}

\usepackage{eso-pic}

\begin{document}
\title{Disturbance-Adaptive Data-Driven Predictive Control: Trading Comfort Violation for Savings in Building Climate Control}
\author{Jicheng Shi, Christophe Salzmann and Colin N. Jones
\thanks{The first author received support from the Swiss National Science Foundation (SNSF) under the NCCR Automation project, grant agreement 51NF40\_180545.}
% \thanks{*: Corresponding author} $^*$
\thanks{Jicheng Shi, Christophe Salzmann and Colin Neil Jones are with the Automatic Control Lab, EPFL, Switzerland.(e-mail: \{jicheng.shi, christophe.salzmann, colin.jones\}@epfl.ch)}\vspace{-1em}
}

% --- overlay notice on the FIRST page only (no layout space consumed)
\AddToShipoutPictureFG*{%
  \AtPageUpperLeft{%
    \raisebox{-1.2cm}{% 纵向位置：数值越大越往下
      \hspace*{1.7cm}% 横向左边距
      \parbox{0.83\paperwidth}{%
    \footnotesize
    \textbf{Accepted manuscript.} Accepted for publication in \emph{IEEE Transactions on Control Systems Technology}, 2026. DOI: \href{https://doi.org/10.1109/TCST.2026.3664129}{\textbf{10.1109/TCST.2026.3664129}}.\par
    © 2026 IEEE. Personal use of this material is permitted. Permission from IEEE must be obtained for all other uses, in any current or future media, including reprinting/republishing this material for advertising or promotional purposes, creating new collective works, for resale or redistribution to servers or lists, or reuse of any copyrighted component of this work in other works.      
      }%
    }%
  }%
}
\maketitle
% \ClearShipoutPictureFG % 避免影响后续页面

%%%%%%%%%%%%%%%%%%%%%%%%%%%%%%%%%%%%%%%%%%%%%%%%%%%%%%%%%%%%%%%%%%%%%%
% acronym
\begin{acronym} %\ac{dr}
\acro{mpc}[MPC]{Model Predictive Control}
\acro{bms}[BMS]{Building Management System}
\acro{dpc}[DPC]{Data-Driven Predictive Control}
\acro{deepc}[DeePC]{Data-enabled Predictive Control}
\acro{dad}[DAD]{Disturbance-Adaptive}
\acro{epfl}[EPFL]{École Polytechnique Fédérale de Lausanne}
\acro{hp}[HP]{Heat Pump}
\acro{hvac}[HVAC]{Heating, Cooling and Ventilation}
\acro{io}[I/O]{Input/Output}
\acro{lti}[LTI]{Linear Time-Invariant}
\acro{mae}[MAE]{Mean Absolute Error}
\acro{qp}[QP]{Quadratic Programming}
\acro{scp}[SCP]{Split Conformal Prediction}

\end{acronym}
%%%%%%%%%%%%%%%%%%%%%%%%%%%%%%%%%%%%%%%%%%%%%%%%%%%%%%%%%%%%%%%%%%%%%%

\begin{abstract} 
Model Predictive Control (MPC) has demonstrated significant potential in improving energy efficiency in building climate control, outperforming traditional controllers commonly used in modern building management systems. Among MPC variants, Data-driven Predictive Control (DPC) offers the advantage of modeling building dynamics directly from data, thereby substantially reducing commissioning efforts. However, inevitable model uncertainties and measurement noise can result in comfort violations, even with dedicated MPC setups. This paper introduces a Disturbance-Adaptive DPC (DAD-DPC) framework that ensures asymptotic satisfaction of predefined violation bounds without knowing the uncertainty and noise distributions. The framework employs a data-driven pipeline based on Willems' Fundamental Lemma and conformal prediction for application in building climate control. The proposed DAD-DPC framework was validated through four building cases using the high-fidelity BOPTEST simulation platform and an occupied campus building, Polydome. DAD-DPC successfully regulated the average comfort violations to meet pre-defined bounds. Notably, the 5\%-violation DAD-DPC setup achieved 30.1\%/11.2\%/27.1\%/20.5\% energy savings compared to default controllers across four cases. These results demonstrate the framework's effectiveness in balancing energy consumption and comfort violations, offering a practical solution for building climate control applications.
\end{abstract}

% For a 5\% violation bound, the framework achieved energy consumption increases of 2.82\%, 2.24\%, 3.75\%, and 5.0\% while reducing comfort violations by 77.62\%, 79.51\%, 73.26\%, and 39.8\%, respectively, compared to other controller configurations.


\section{Introduction}

The building sector accounts for approximately 34\% of global energy consumption, with a significant portion attributed to Heating, Ventilation, and Air Conditioning (HVAC) systems~\cite{united2023}. These systems are typically managed by Building
Management Systems (BMS) using classical control strategies, such as rule-based controllers~\cite{salsbury2005survey}. However, these traditional approaches often fall short in efficiency, presenting a clear opportunity to improve energy performance while maintaining occupant comfort.

Model Predictive Control (MPC) has emerged as a promising approach to address this challenge~\cite{drgovna2020all,killian2016ten}. MPC leverages a predictive building model to optimize economic objectives while satisfying comfort constraints, with control actions applied in a receding-horizon manner. Compared to traditional methods, MPC demonstrates superior control performance and robustness in building climate control~\cite{balali2023energy,stoffel2023evaluation,pergantis2024field}. Nonetheless, implementing MPC faces many practical challenges, particularly in accurately modeling the complex dynamics of buildings~\cite{drgovna2020all,killian2016ten}.

To mitigate this modeling challenge, \ac{dpc} using data-driven or gray-box techniques has gained considerable attention~\cite{xiao2023building,stoffel2023evaluation,bunning2022physics}. Compared to MPC with physics-based models~\cite{jorissen2019taco}, DPC requires less engineering effort and expert knowledge. For example, \cite{lian2023adaptive} demonstrated that a linear DPC, developed using only two days of operational building data, achieved an 18\% reduction in energy consumption compared to an industrial control method in a real-world application. Furthermore, \cite{di2022lessons} highlighted the potential of different data-driven modeling methods for addressing specific building control scenarios through a pragmatic, qualitative comparison study. 
Despite these advances, both physics-based and data-driven approaches remain susceptible to model uncertainties and measurement noise, often resulting in suboptimal performance or comfort violations~\cite{drgovna2020all}.

To address uncertainties, robust MPC approaches have been proposed to ensure constraint satisfaction~\cite{bemporad2007robust}.
\change{Recent work combines robust MPC with data-driven methods to learn system dynamics and preserve robust constraint guarantees~\cite{alanwar2022robust,de2024koopman,dubied2025robust,schimperna2024robust}. For instance, \cite{alanwar2022robust} proposes a data-driven robust MPC for unknown linear systems that constructs reachable sets directly from noisy input–output data, guaranteeing robust constraint satisfaction under bounded process and measurement noise. In building control, robust MPC has been integrated with data-driven and gray-box models identified from data, including resistor-capacitor (RC) models~\cite{gao2023energy}, Gaussian processes~\cite{maddalena2022experimental}, and neural networks~\cite{li2022tube,mahmood2023data}. For example, \cite{gao2023energy} identifies a gray-box RC model and designs a tube-based MPC that tightens constraints, reducing operating cost by at least 24\% and improving the temperature tracking under multiple uncertain predictions. 
In a complementary direction, \cite{hu2023multi} constructs a data-driven, clustered weather-forecast uncertainty set and embeds it in an affine-disturbance-feedback robust MPC, yielding up to 8.8\% energy savings compared to conventional robust MPC approaches.}
However, their reliance on worst-case optimization \change{can} lead to excessive conservatism and higher costs. 

In contrast, stochastic MPC offers a more balanced approach by allowing occasional comfort violations~\cite{mesbah2016stochastic}, utilizing chance constraints to meet comfort requirements with a predefined probability. For example, \cite{oldewurtel2012use} demonstrated the application of applied stochastic MPC in various building cases, highlighting its capability to balance energy efficiency and occupant comfort through sensitivity analysis of probability parameters. Nevertheless, most stochastic MPC methods are still conservative due to sufficient conditional chance constraints\cite{korda2012stochastic} and depend on precise knowledge of disturbance distributions, which is rarely feasible in real-world applications~\cite{mesbah2016stochastic}. This raises the critical question of how to effectively manage comfort violations to achieve cost savings without exact disturbance characterization.

Recent efforts have explored stochastic MPC formulations that enforce average constraint violation bounds over time~\cite{korda2012stochastic,korda2014stochastic,fleming2017time,shi2024dad,oldewurtel2013adaptively,ghosh2025adaptive}. These methods adapt disturbance bounds or constraint tightening based on closed-loop violations, directly related to the comfort criteria in building climate control. \change{The authors of}~\cite{korda2012stochastic} demonstrated that such approaches could reduce control costs compared to traditional stochastic MPC. More recently, \cite{shi2024dad} proposed a disturbance-adaptive scheme that ensures asymptotic and robust average violation bounds, even with inaccurate disturbance quantification. 
% However, practical gaps remain in applying these methods to building climate control, as they typically rely on linear state-space models and full state measurements, which are often impractical in real-world scenarios.
However, \change{applying existing average violation guarantees to building climate control is challenging because the guarantees rely on assumptions that are difficult to satisfy or implement in buildings. In particular, several results (\!\cite{korda2012stochastic,korda2014stochastic,fleming2017time,shi2024dad}) assume the existence of a robust invariant set computed from a state-space model. Because buildings are modeled in various ways (such as adaptive data-driven predictors) and are driven by time-varying exogenous inputs (such as weather and occupancy), computing such sets without excessive conservatism can be challenging. Other results (\!\cite{oldewurtel2013adaptively,ghosh2025adaptive}) assume a control policy with a one-step corrective property: if the average violation so far is too large or too small, the next-step violation will certainly disappear or occur. This is a strong assumption under actuator saturation, duty-cycling, and the large thermal inertia and slow thermal dynamics of buildings.}
Furthermore, while some studies have explored building control applications~\cite{oldewurtel2013adaptively,korda2014stochastic}, these have primarily used simple linear time-invariant models, lacking validation in high-fidelity simulations or real-world experiments.

This paper extends the disturbance adaptive method~\cite{shi2024dad} to a DPC framework \change{without computing a robust invariant set}, enabling a trade-off between the average comfort violations and cost savings in its application of building climate control. The key contributions are as follows:

\begin{itemize}
    \item[1)] Disturbance-Adaptive DPC (DAD-DPC) Framework: A novel DAD-DPC framework is introduced, providing an asymptotic guarantee on average constraint violations without knowing the disturbance distribution. The guarantee is not limited to any specific predictive model.
    \item[2)] Efficient Design for Building Climate Control: A tailored DAD-DPC design method is developed, leveraging Willems' Fundamental Lemma and conformal prediction. The approach enables a data-driven workflow with low commissioning effort and results in a computationally efficient formulation.
    \item[3)] Comprehensive Validation: The proposed framework was validated across four building cases using a high-fidelity simulation platform and an occupied campus building. The closed-loop average comfort violations precisely met predefined bounds. To the best of the authors’ knowledge, this study represents the first experimental demonstration of achieving this goal in building climate control.  
    \item[4)] Trade-off Between Comfort Violations and Energy Consumption: Compared to the conservative default controller in four cases, DAD-DPC with 5\% average violations, suggested by a European standard, reduced energy consumption by 30.1\%/11.2\%/27.1\%/20.5\%, respectively. Furthermore, compared to more relaxed violation bounds, the 5\% DAD-DPC setting increased energy use by only 2.8\%/3.7\%/3.7\%/5.0\% while reducing violations by 77.6\%/79.6\%/73.2\%/39.8\%.
\end{itemize}

% Comprehensive Validation: The proposed framework was validated using a high-fidelity simulation platform and an actual campus building testbed, considering four different building cases and a series of different violation targets.
%     \item[4)] Trading comfort violations for savings: 
%  The results demonstrated that the closed-loop average comfort violations precisely met predefined bounds. To the best of the authors' knowledge, this study represents the first experimental demonstration of achieving this goal in building climate control.   
%     Compared to the conservative default controller in four Cases, DAD-DPC with 5\% average violation, suggested by a European standard, reduced 30.13\%/12.50\%/27.07\%/20.5\% energy consumption, respectively. 
% Therein, the proposed DAD-DPC with the 5\% violation bound increased 4.28\%/1.33\%/4.74\%/5.0\% energy consumption with 77.64\%/80.40\%/77.40\%/39.8\% fewer violations compared to their counterparts.
% , BOPTEST~\cite{blum2021building},

% We propose several practical extensions of the approach, including the computationally efficient formulation of DAD-DPC and various constraint violation bounds.

% add:
% - results of trading (main value), and comparison to default controllers: proving that the framework brings values
% - reference

\section{PRELIMINARIES}

This section introduces the notations and foundational concepts used in the paper and describes the building climate control problem.

\vspace{-0.3em}
\subsection{Notation}
The identity matrix of size n is denoted as $I_n\in\mathbb{R}^{n\times n}$. The sets of non-negative and positive integers are represented by $\N$ and $\N_{+}$, respectively. For a range of consecutive integers from $i$ to $j$, we use $\N_i^j = \{i, i + 1, \dots, j\}$. For a vector $v$, $v_t$ represents its value at time $t$, 
$v^{(i)}$ denotes its $i$-th element in $v$, and $v_{i|t}$ refers to the $i$-step prediction from time $t$.
Additionally, $\{v_t\}_{t=i}^j$ denotes the sequence of $v_t$ from time $i$ to $j$. In a controlled system, $u, y, w$ represent the control inputs, outputs, and external inputs, respectively.

% The $i$-th element in $v$ is denoted by $v^{(i)}$, and the $i$-th row in a matrix $M$ is denoted by $M(i,:)$

% $z_t, u_t, y_t, d_t, e_t$ respectively denote the states, inputs, outputs, external disturbances, and measurement noise in a controlled system at time t.
% $\mathbf{0}_{n} \in\mathbb{R}^{n}$ denotes a zero vector.

% \noindent\textbf{Notation:} $\mathcal{N}(\mu,\Sigma)$ denotes a Gaussian distribution with mean $\mu$ and covariance $\Sigma$. $\otimes$ is the symbol for the Kronecker product. $\times$ is the symbol for the set product operation \change{(i.e. $A\times B:= \{(x,y)|x\in A,\;y\in B\}$)}, $\oplus$ is the symbol of Minkowski sum \change{(i.e. $A\oplus B:=\{x|\exists\;x_1\in A,\;x_2\in B,\; x=x_1+x_2\}$)} and $\ominus$ is the symbol of \change{Pontryagin} difference \change{(i.e. $A\ominus B :=\{x|x+y\in A,\;\forall\;y\in B\}$)}. $\text{colspan}(A)$ denotes the column space (i.e. range) of the matrix A. $I_n$ is the $n\times n$ identity matrix, \change{$\textbf{O}$ and $\mathbf{0}$ are zero matrix and zero vector respectively}. $\lVert \cdot\rVert$ denotes the Euclidean two-norm. $x: = \{x_i\}_{i=1}^T$ denotes a sequence of size $T$ indexed by $i$. $x_i$ denotes the measurement of $x$ at time $i$, and $x_{1:L}:=[x_1^\top,x_2^\top\dots x_L^\top]^\top$ denotes a concatenated sequence of $x_i$ ranging from $x_1$ to $x_L$, and we drop the index to improve clarity if the intention is clear from the context.

\vspace{-0.3em}
\subsection{Problem Statement} 
This paper focuses on building climate control, aiming to enhance the energy efficiency of HVAC systems while maintaining comfortable indoor temperature levels~\cite{oldewurtel2012use}.

Consider a building system where the outputs $y$ represent the indoor temperature, the control inputs $u$ correspond to the operational variables of HVAC components, and the external inputs $w$ account for factors such as outdoor temperature and solar radiation. The objective cost at time $t$ is denoted by $I(y_t,u_t)$. For example, if $u_t$ corresponds to the power consumption of a \ac{hp},  the cost can be expressed as $I(y_t,u_t)= p_t u_t$, where $p_t$ is the current electricity price. 
The control inputs must satisfy system constraints, $u_t \in \mathcal{U}$, as dictated by the specifications of the HVAC system. 

The indoor temperature should satisfy time-varying polyhedral comfort constraints defined as 
\begin{align*}
   \mathcal{Y}_t = \{y|F_{y,t} \; y \leq f_{y,t} \} \; . 
\end{align*}
For example,  a single-zone office may use the following box constraints:
\begin{align} \label{eqn:prob_ycons}
    \mathcal{Y}_t= \begin{cases}
       \{y|{21}^\circ C \leq y 
 \leq {25}^\circ C \}, &  \text{from } 8 \text{ a.m. to } 6 \text{ p.m.} \\
        \{y|{19}^\circ C \leq y 
 \leq {27}^\circ C \}, & \text{otherwise.}
    \end{cases}
\end{align}
These constraints relax the temperature range during non-working hours to save energy. Notably, occasional violations of comfort constraints are permitted under certain building regulations~\cite{comite2007indoor,sourbron2011evaluation}. To account for this, \change{and as a central focus of this paper,} we formalize an asymptotic average comfort violation \change{bound} as follows:
\begin{align} \label{eqn:cons_vio_asy}
    \lim_{t\rightarrow\infty} \frac{\sum_{i=1}^{t} v_i}{t} \leq  \alpha \; ,
\end{align}
where $v_t$ is a binary variable indicating whether a comfort violation occurs at time $t$: 
\begin{align*}
    v_t = \begin{cases}
        0 \; , & \text{if } y_t \in \mathcal{Y}_t \; ; \\
        1 \; , & \text{if } y_t \notin \mathcal{Y}_t \; .
    \end{cases}
\end{align*}

In the constraint~\eqref{eqn:cons_vio_asy}, $\alpha \in (0, 1]$ specifies the allowable average violation bound. According to \cite{sourbron2011evaluation}, the value of $\alpha$ should be determined through pre-design discussions between building designers and clients. As shown in~\cite{oldewurtel2012use}, increasing $\alpha$ results in higher comfort violations for greater energy savings, which was validated using a stochastic MPC method. However, the approach relied on sensitivity analysis of a probability parameter to \change{satisfy} the desired $\alpha$-level \change{bound for} average violations in the closed loop~\cite{oldewurtel2012use}. This limitation is common in stochastic MPC methods, where fixed conditional chance constraints introduce excessive conservatism and practical estimates of disturbance distributions are often quite inaccurate.

\begin{figure}[t]
    \centering
    \includegraphics[width=1.0\linewidth]{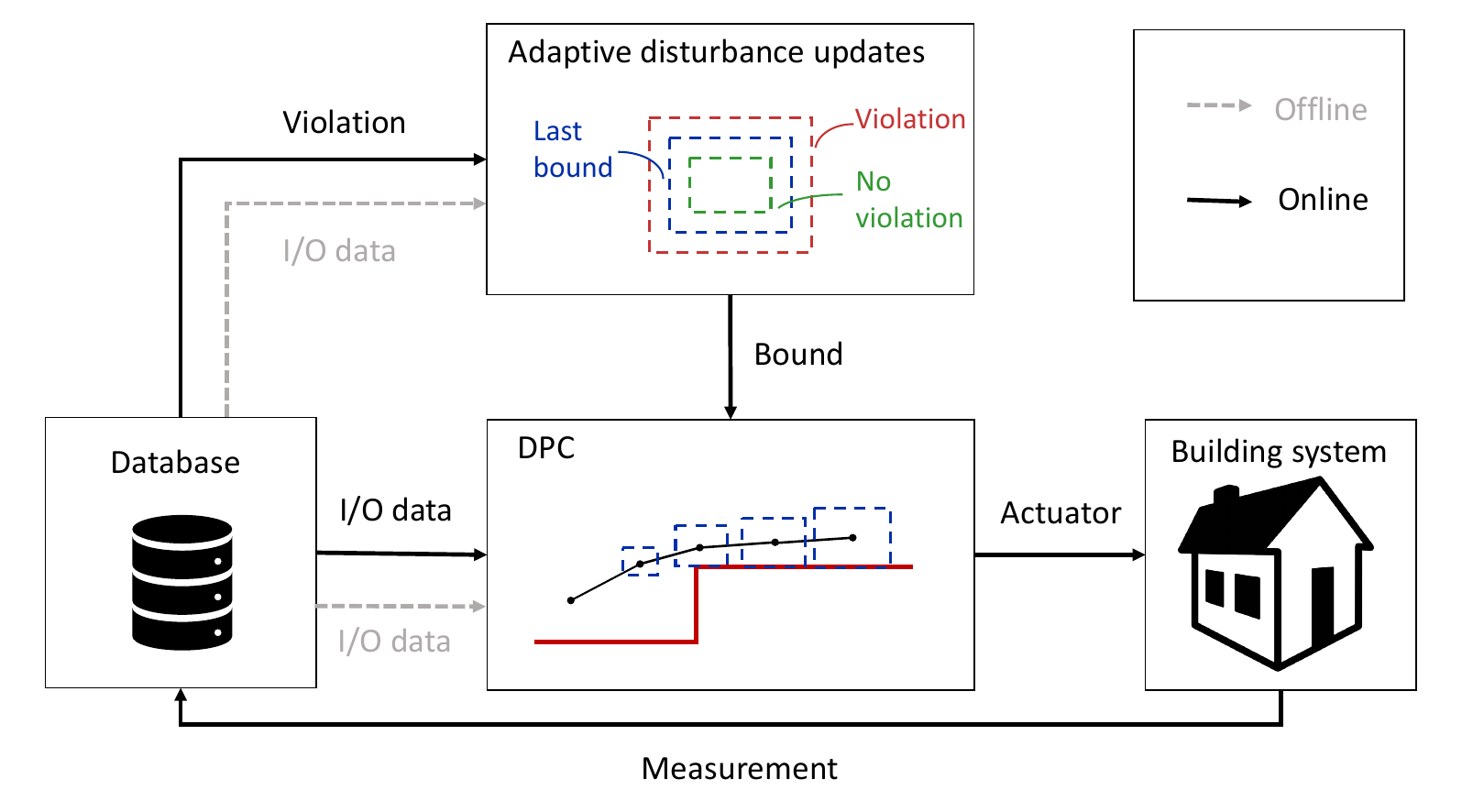}
        \caption{Illustration of the DAD-DPC. Offline phase: Historical \ac{io} data are used to construct a disturbance bound estimator and a DPC controller. Online phase: The disturbance bound is adaptively updated based on the current violation condition, and the DPC then utilizes the updated bound and the latest \ac{io} measurement to compute the optimal input, which is applied to the building system.}
    \label{fig:dad_concept}
    \vspace{1em}
\end{figure}

\section{Methodology}

This paper proposes a novel DAD-DPC framework designed to meet the closed-loop violation constraint~\eqref{eqn:cons_vio_asy}. Its overall structure is illustrated in Figure~\ref{fig:dad_concept}, with building climate control as the application example. 
Unlike traditional stochastic or robust MPC methods that rely on fixed disturbance bounds or constraint tightening, the DAD-DPC framework adaptively updates the disturbance bound based on the observed closed-loop violations.

The methodology is presented in four parts. Section~\ref{sect:method_daddpc} introduces the overall DAD-DPC framework, detailing its \change{three} main components, and specifies a key property required for theoretical guarantees.
Section~\ref{sect:method_guarantee} establishes that this property is sufficient to guarantee the violation constraint~\eqref{eqn:cons_vio_asy}.
In addition, Section~\ref{sect:III_heuristic} presents an informal property that achieves a heuristic feedback mechanism for violation regulation in DAD-DPC.
Finally, Section~\ref{sect:method_design} provides one feasible design of DAD-DPC tailored for building climate control.
This implementation satisfies the required properties and enables a data-driven workflow with low commissioning effort.

% \vspace{-0.3em}
\subsection{Disturbance-Adaptive Data-Driven Predictive Control} \label{sect:method_daddpc}

The DAD-DPC framework comprises \change{three} main components:
\begin{itemize}
    \item A disturbance bound estimator, denoted by $\mathcal{D}(\sigma)$, \change{with $\sigma \!\! \in \!\!(0,1]$.}  It outputs a bounded set in which the disturbance is expected to lie. The size is defined in terms of set inclusion: a larger bound contains more possible disturbance values.
    \item A DPC controller, denoted by \changeC{$\pi(\cdot, \mathcal{D}(\sigma))$}.
    \item \changeC{A backup controller $\pi^B$ (a feedback law mapping measurements to inputs)} \change{and a designer-chosen set $\mathcal{Y}_{\text{lim}} \supseteq \mathcal{Y}_t$. The formal requirements on $\pi^B$ and $\mathcal{Y}_{\text{lim}}$ are stated in Property 1.}
\end{itemize}
$\sigma \in [0,1]$ is a parameter that controls the conservatism of the bound. \change{We denote $z_t$ as the initialization variable used to start the model prediction} in the DPC controller $\pi(z_t, \mathcal{D}(\sigma))$. \change{The definition of $z_t$ depends on the predictive model: for example, it may be the state in a state-space model, estimated states obtained from an observer, or the most recent measured outputs and inputs in an autoregressive exogenous model.} 
\changeC{$z_t$ may also include the output of a forecaster, such as the weather forecast in building control.}
\change{The proposed framework does not require full-state feedback but only the variables necessary to initialize the predictive model.} 
% \change{In step 1), predictable external inputs are obtained, such as the weather forecast in building control.}

% For any $\mathcal{D}(\sigma)$ and $\pi(z_t, \mathcal{D}(\sigma))$ satisfying Property 1, 
The online operation of the DAD-DPC framework for a target violation bound $\alpha$ in~\eqref{eqn:cons_vio_asy} is outlined in Algorithm~\ref{alg:dad_dpc}. The additional parameters $\eta$, $\alpha_t$ are defined within the algorithm.

\begin{algorithm} 
  \caption{DAD-DPC: Online Operation} \label{alg:dad_dpc}
{
\textbf{Input:} Target violation bound $\alpha \in (0,1]$, disturbance bound estimator $\mathcal{D}(\sigma)$,  control policy $\pi(z_t, \mathcal{D}(\sigma))$, initial value $\alpha_0 \change{\geq 0}$ of a variable $\alpha_t \in \R$, update rate $\eta > 0$, an output set $\mathcal{Y}_{\text{lim}} \supseteq \mathcal{Y}$, \change{last output $y_{-1}\in\mathcal{Y}_{\text{lim}}$}.
\begin{itemize} [nolistsep,leftmargin=1.5em]
    \item[1)] At time $t$, obtain the current output $y_t$ and state $z_t$.
    \item[2)] \changeC{Compute the violation indicator $v_t =\begin{cases}
        0 \; , & \text{if } y_t \in \mathcal{Y}_t \; ; \\
        1 \; , & \text{if } y_t \notin \mathcal{Y}_t \; .
    \end{cases}$}
Update $\alpha_t$ based on $v_t$ and compute its truncated value $\bar{\alpha}_t \in [0,1]$: 
    {\setlength{\abovedisplayskip}{1.6pt}\setlength{\belowdisplayskip}{1.6pt}
    \begin{equation}
    \begin{aligned} \label{eqn:dad_alpha}
        & \alpha_{t} = \alpha_{t-1} + \eta(\alpha - v_t) \; , \\
        & \bar{\alpha}_t = \min(\max(\alpha_{t},0),1) \; .
    \end{aligned}
    \end{equation}
    }       
    \item[3)] Apply the input according to \change{the condition:} 
    {\setlength{\abovedisplayskip}{1.6pt}\setlength{\belowdisplayskip}{1.6pt}
    \begin{align} \label{eqn:dad_mpc_u}
    \change{u_t =
    \begin{cases}
        \pi(z_t,  \mathcal{D}(\bar{\alpha}_t)) \; , & \text{if } y_t \in \mathcal{Y}_{\text{lim}}  \; \text{and} \; \bar{\alpha}_t > 0 \; ; \\
        \pi^B \; , & \text{if } y_t \notin \mathcal{Y}_{\text{lim}} \; \text{or}  \; \bar{\alpha}_t = 0 \; . \\
    \end{cases}}
    \end{align}
    }      
    \item[4)] Wait until the next sampling time, update $t\leftarrow t+1$ and repeat from step 1).    
\end{itemize}
}
\end{algorithm}

Algorithm~\ref{alg:dad_dpc} adaptively updates $\alpha_t$, which dynamically adjusts the disturbance bound used in the control policy, as illustrated in Figure~\ref{fig:dad_concept}. Specifically, as shown in~\eqref{eqn:dad_alpha}, $\alpha_t$ is initialized with $\alpha_0$ and updated based on the violation condition $v_t$, and the parameter $\eta$ determines the time constant. The truncated $\bar{\alpha}_t$ determines the time-varying disturbance bound $ \mathcal{D}(\bar{\alpha}_t)$. \change{As defined in the switching rule~\eqref{eqn:dad_mpc_u}, the control policy applies either the DPC controller $\pi(z_t, \mathcal{D}(\bar{\alpha}_t))$ or the backup controller $\pi^B$ depending on whether $y_t\in \mathcal{Y}_{\text{lim}}$ and $\bar{\alpha}_t>0$}.  
This adaptive update of the disturbance bound distinguishes the DAD-MPC from traditional stochastic MPC methods, which typically employ fixed constraint-tightening values. \change{
The effect of the adaptive term and of $\pi(z_t,\mathcal{D}(\sigma))$ is discussed in Section~\ref{sect:III_heuristic}.}

Importantly, DAD-DPC is not limited to specific predictive models or disturbance bound estimation techniques. 
\change{Instead, the backup controller $\pi^B$ and the set $\mathcal{Y}_{\text{lim}}$ are required to satisfy the following property:
\begin{itemize}
    \item \textbf{Property 1} (Average violation under backup controller)  There exist finite constants $\bar{\Delta}\in\mathbb{N}_+$ and $\epsilon \in [0,\alpha]$ such that:
for any time $t_0$ and any $\Delta\in\mathbb{N}_+$ with $\Delta\ge \bar{\Delta}$ and
$u_{t_0+i}=\pi^B$ for all $i=0,1,\dots,\Delta-1$,
    if $y_{t_0-1}\in\mathcal{Y}_{\text{lim}}$, then we have $\frac{1}{\Delta}\sum_{i=1}^{\Delta} v_{t_0+i} \ \le\ \alpha - \epsilon$.
\end{itemize}}

\begin{figure}[t]
    \centering
    \includegraphics[width=1.0\linewidth]{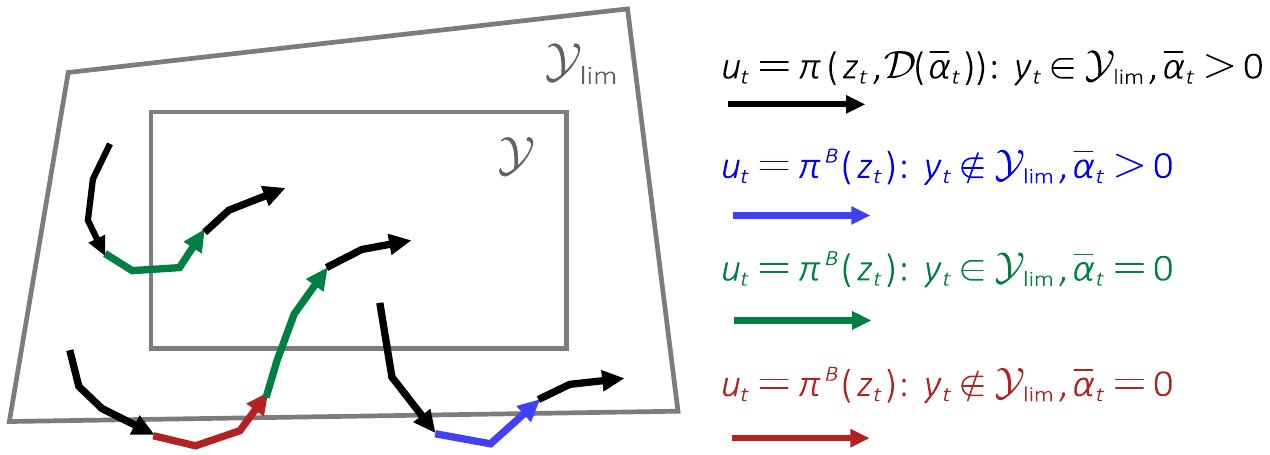}
        \caption{\change{Illustration of the switching rule~\eqref{eqn:dad_mpc_u} in Algorithm~\ref{alg:dad_dpc}. Colored segments mark which input is activated along three example trajectories.} 
        \changeC{The backup controller $\pi^B$ is activated either when $y_t$ remains within $\mathcal{Y}_{\text{lim}}$ but the constraint is violated for too long such that $\bar{\alpha}_t = 0$, or when $y_t$ leaves $\mathcal{Y}_{\text{lim}}$.
        The set $\mathcal{Y}_{\text{lim}} \supseteq \mathcal{Y}_t$ defines the precondition under which $\pi^B$ operates according to Property 1, and its design methods are discussed in Remark~\ref{remark:assum2&3}.}}
    \label{fig:Y_lim}
    % \vspace{1em}
\end{figure}

\change{Property 1 states that, whenever its precondition holds (i.e., $y_{t_0-1}$ is inside $\mathcal{Y}_{\text{lim}}$), any sufficiently long interval on which  $\pi^B$ is applied from time $t_0$ keeps an average violation not exceeding $\alpha-\epsilon$. Under Algorithm~\ref{alg:dad_dpc}, this precondition is met at each switch from  $\pi(z_t, \mathcal{D}(\sigma))$ to $\pi^B$ according to~\eqref{eqn:dad_mpc_u}, as illustrated in Figure~\ref{fig:Y_lim}. In the next section, Theorem~\ref{thm:bound_suf_2} shows a formal guarantee that Algorithm~\ref{alg:dad_dpc} and Property 1 ensure satisfaction of the bound~\eqref{eqn:cons_vio_asy}. In addition, if $\epsilon>0$ in Property 1, any activation of $\pi^B$ lasts only for a finite number of steps, after which the policy switches back to the DPC controller $\pi(z_t, \mathcal{D}(\sigma))$.}

\begin{remark} \label{remark:assum2&3}
\change{In the DAD-DPC framework}, Property 1 only requires the existence of finite constants \change{$\bar{\Delta}$ and $\epsilon$}, without needing their explicit values.
\change{This gives flexibility in choosing  $\pi^B$ and $\mathcal{Y}_{lim}$.}\\
\changeC{\textbf{(i)} One option is to design $\pi^B$ to practically stabilize the system to a constraint-admissible set. Choose a compact subset of its domain of attraction, and define $\mathcal{Y}_{\text{lim}}$ as a robust backward reachable set of this subset.}
\change{For linear systems, \cite{shi2024dad} and~\cite{korda2012stochastic} present a systematic approach based on robust control invariant sets. This design ensures that whenever $\pi^B$ is activated, constraint violations vanish after a user-defined finite number of steps, thereby satisfying Property~1.} \\
\change{\textbf{(ii)} For building systems,} Property 1 is generally non-restrictive, because it aligns with typical HVAC system capabilities, which are designed to provide sufficient heating and cooling capacity for effective temperature regulation. \change{Therefore, we can design $\pi^B$ by industrial default controllers and $\mathcal{Y}_{\text{lim}}$ within the effective operating range based on engineering knowledge. Examples are shown in Sections IV and V.}
\end{remark}

\subsection{Guarantees by DAD-DPC} \label{sect:method_guarantee}

% This section will provide sufficient conditions under which the system controlled by the DAD-DPC framework satisfies the asymptotic violation bound~\eqref{eqn:cons_vio_asy}, as formalized in  Lemma~\ref{thm:bound_suf_1} and Theorem~\ref{thm:bound_suf_2}.

This section provides the guarantee of the asymptotic violation bound~\eqref{eqn:cons_vio_asy} for a system controlled by the DAD-DPC framework under Property 1. We begin by presenting sufficient conditions, formalized in Lemma~\ref{thm:dad_bound} and~\ref{thm:bound_suf_1}. These results form the basis for Theorem~\ref{thm:bound_suf_2}, which provides the final guarantee.

First, we establish that at time $t$, the closed-loop average violation under the DAD-MPC is bounded related to the minimal and maximal $\alpha_{t}$ from initial time, denoted as $\alpha_{min,t}:=\min_{i \in \N_0^t} \alpha_i, \alpha_{max,t}:=\max_{i \in \N_0^t} \alpha_i$, respectively~\cite{shi2024dad}. 

\begin{lemma} \label{thm:dad_bound}
    Control a system by DAD-DPC following Algorithm~\eqref{alg:dad_dpc}. The average violation at time $t$ is bounded as:
    \begin{align} \label{eqn:thm_bound}
    \alpha +  \frac{\alpha_0-\alpha_{max,t}}{t\eta} \leq \frac{\sum_{i=1}^{t} v_i}{t} \leq \alpha +  \frac{\alpha_0-\alpha_{min,t}}{t\eta} \; .
    \end{align}
\end{lemma}

\begin{proof}
    By recursively expanding the update equation~\eqref{eqn:dad_alpha} , we derive:
    \begin{align*}
        \alpha_{t} = \alpha_{0} + \eta\sum_{i=1}^{t}(\alpha - v_i) \; .
    \end{align*}
    Since \change{$\alpha_{min,t} \leq \alpha_t \leq \alpha_{max,t}$}, $\eta > 0$ and $t \geq 1$, the bounds~\eqref{eqn:thm_bound} follow directly.  
\end{proof}

Based on Lemma~\ref{thm:dad_bound}, the existence of a lower bound for $\alpha_t$, provides the first sufficient condition for ensuring~\eqref{eqn:cons_vio_asy}. Subsequently, three other violation bounds are introduced with their sufficient conditions.

\begin{lemma} \label{thm:bound_suf_1}
    Control a system by DAD-DPC following Algorithm~\eqref{alg:dad_dpc}. If $\exists \mathrm{A}\in \R$ such that $\alpha_t \geq \mathrm{A}, \forall t \in \N_{+}$, the bound~\eqref{eqn:cons_vio_asy} is satisfied. Furthermore, if $A \geq \alpha_0$,  a stricter bound,
    \begin{align} \label{eqn:cons_vio_rob}
        \frac{\sum_{i=1}^{t} v_i}{t} \leq  \alpha, \forall t \in \N_{+} \; ,
    \end{align} holds. Similarly, if $\exists \mathrm{B}\in \R$ such that $\alpha_t \leq \mathrm{B}, \forall t \in \N_{+}$, another asymptotic lower bound is satisfied:
    \begin{align} \label{eqn:cons_vio_asy_lower}
        \lim_{t\rightarrow\infty} \frac{\sum_{i=1}^{t} v_i}{t} \geq  \alpha \; .
    \end{align}
\end{lemma}

\begin{proof}
If $\alpha_t \geq \mathrm{A}$, then $\alpha_{min,t} \geq A$. Therefore, by adding infinite limits to both sides of the inequality~\eqref{eqn:thm_bound}:
    \begin{align*}
         \lim_{t\rightarrow\infty}     \frac{\sum_{i=1}^{t} v_i}{t} &\leq   \alpha +  \lim_{t\rightarrow\infty} \frac{\alpha_0-\alpha_{min, t}}{t\eta} \\
         &\leq \alpha +  \lim_{t\rightarrow\infty} \frac{\alpha_0-A}{t\eta} \\
         & = \alpha \; .
    \end{align*}  
    The equality holds because $\alpha_0$ and $\mathrm{A}$ are finite constants. 
    If additionally $A \geq \alpha_0$, then $\alpha_{min,t} \geq A\geq \alpha_0$, and~\eqref{eqn:thm_bound} immediately implies the stricter bound~\eqref{eqn:cons_vio_rob} by:
    \begin{align*}
        \frac{\sum_{i=1}^{t} v_i}{t} \leq   \alpha +  \frac{\alpha_0-\alpha_{min}}{t\eta} \leq \alpha \; .
    \end{align*}    
    \vspace{-0.5em}
    \changeC{Similarly, the asymptotic lower bound~\eqref{eqn:cons_vio_asy_lower} can be proven.}
\end{proof}

% While Lemma~\ref{thm:bound_suf_1} provides the sufficient condition for~\eqref{eqn:cons_vio_asy}, it does not specify the direct requirements on the design of $\pi(z_t, \mathcal{D}(\alpha_t))$ and $\mathcal{D}(\alpha_t)$. 
%  To address this, we introduce a more practical sufficient condition:
 
Finally, Theorem~\ref{thm:bound_suf_2} is derived based on Lemma~\ref{thm:bound_suf_1}, assuming Property 1 holds.

\begin{theorem} \label{thm:bound_suf_2}
    Control a system by DAD-DPC following Algorithm~\eqref{alg:dad_dpc}. If $\alpha_0\ge 0$, $y_{\change{-1}}\in\mathcal{Y}_{\text{lim}}$, and Property~1 holds, then the following lower bound for $\alpha_t$ holds:
    \begin{align*}
        \alpha_t \geq \mathrm{A}:= -\eta(1-\alpha)(\bar{\Delta}+1) \; , \; \forall t \geq 0 \; .
    \end{align*}
    Consequently, the asymptotic violation bound~\eqref{eqn:cons_vio_asy} holds by  Lemma~\ref{thm:bound_suf_1}. \change{If, in addition, $\epsilon > 0$ in Property~1, then any activation of $u_t=\pi^B$ lasts for a finite number of steps.}
\end{theorem}

\begin{proof}
{\color{black}
By~\eqref{eqn:dad_mpc_u}, if $u_t=\pi(z_t,\mathcal{D}(\bar{\alpha}_t))$, then $\bar{\alpha}_t>0$ and hence $\alpha_t>0 > \mathrm{A}$; thus it suffices to bound $\alpha_t$ when $u_t=\pi^B$.

Let $[\tau,\tau+\Delta-1]$ with $\Delta\in\mathbb{N}_+\cup\{\infty\}$ be \changeC{any interval during which $u_t=\pi^B$ and $u_{\tau-1} = \pi(z_t,\mathcal{D}(\bar{\alpha}_t))$ if $\tau\neq0$.}
At its start, \changeC{we have $\alpha_\tau \geq -\eta(1-\alpha)> \mathrm{A}$ because:}
\begin{itemize}[leftmargin=1.2em,nosep]
\item[-] If $\tau=0$, then $y_{-1}\in\mathcal{Y}_{\text{lim}}$ and $\alpha_\tau=\alpha_0\ge 0\geq -\eta(1-\alpha)$ by assumption.
\item[-] If $\tau>0$, then by the switching rule, $y_{\tau-1}\in\mathcal{Y}_{\text{lim}}$ and $\bar{\alpha}_{\tau-1}>0$, so $\alpha_{\tau-1}>0$. Then as $v_{\tau} \in \{0,1\}$, we have
$\alpha_{\tau}=\alpha_{\tau-1}+\eta(\alpha-v_{\tau}) \geq -\eta(1-\alpha)$.
\end{itemize}

We then analyze $\alpha_{\tau+m}$ for any integer $m$ with $1\le m\le \Delta-1$. Expanding~\eqref{eqn:dad_alpha} gives
\begin{align*}
\alpha_{\tau+m} \;=\; \alpha_{\tau} + \eta \sum_{i=1}^{m}(\alpha - v_{\tau+i}).    
\end{align*}
We proceed by considering two cases of $m$.

(i) Case $m\le \bar{\Delta}$.

Since \changeC{$v_{\tau+i} \in \{0,1\}$}, we obtain
\begin{align*}
\alpha_{\tau+m}  & \geq -\eta(1-\alpha) - \eta(1-\alpha)m
 = -\eta(1-\alpha)(m+1) \\
 & \geq -\eta(1-\alpha)(\bar{\Delta}+1)
 = \mathrm{A}.
\end{align*}

(ii) Case $m\ge \bar{\Delta}$.

Because $y_{\tau-1}\in\mathcal{Y}_{\text{lim}}$ and $u_{\tau+i}=\pi^B$ for $i=0,\dots,m-1$, Property~1 implies
$\sum_{i=1}^{m} v_{\tau+i} \le m(\alpha-\epsilon)$, hence
\begin{align*}
 \alpha_{\tau+m}   & \geq -\eta(1-\alpha) + \eta m\,\epsilon \\
 &\geq  -\eta(1-\alpha)(\bar{\Delta}+1)
 = \mathrm{A}.    
\end{align*}

Thus $\alpha_t\ge \mathrm{A}$ for all $t\ge 0$.} \changeC{The bound~\eqref{eqn:cons_vio_asy} holds by  Lemma~\ref{thm:bound_suf_1}.}

\changeC{We then prove that when $\epsilon > 0$, every backup interval $[\tau,\tau+\Delta-1]$ is finite.}
\changeC{Suppose, by contradiction, that there exists an interval with $\Delta = \infty$.}
\change{For any $m$ with $m\geq \bar{\Delta}$, Property 1 implies  $\frac{1}{m}\sum_{i=1}^{m} v_{\tau+i} \leq \alpha - \epsilon$ and we have
$\alpha_{\tau+m}\geq -\eta(1-\alpha)+\eta m\,\epsilon$, as discussed in (ii).
Let
\begin{align*}
    t_1:=\max\!\Big\{\bar{\Delta},\ \big\lceil\tfrac{1-\alpha}{\epsilon}\big\rceil+1\Big\}.
\end{align*}
Then for all $m\geq t_1$, we have $\alpha_{\tau+m}\geq \eta\epsilon>0$.}
\changeC{Moreover, because $\frac{1}{m}\sum_{i=1}^{m} v_{\tau+i} \leq \alpha-\epsilon < 1$ and $v_{\tau+i} \in \{0,1\}$, there must exist a finite $j\geq t_1$ such that $v_{\tau+j}=0$ and $y_{\tau+j}\in\mathcal{Y}_{\tau+j}\subseteq\mathcal{Y}_{\text{lim}}$. At this point, the switching rule~\eqref{eqn:dad_mpc_u} enforces $u_{\tau+j}=\pi(z_t,\mathcal{D}(\bar{\alpha}_t))$, contradicting $\Delta = \infty$. Therefore, every such interval is finite when $\epsilon>0$.}
\end{proof}

\subsection{Heuristic property for practical implementation} \label{sect:III_heuristic}

% \changeT{Once Property 1  ensures satisfaction of the asymptotic violation bound~\eqref{eqn:cons_vio_asy}, the design of the full controller remains flexible.} 
Theorem~\ref{thm:bound_suf_2} establishes that Property 1  is sufficient to ensure satisfaction of the asymptotic average violation bound~\eqref{eqn:cons_vio_asy}, regardless of the specific choice of $\pi(z_t, \mathcal{D}(\sigma))$ for $\sigma > 0$ and $\mathcal{D}(\sigma)$.
However, the closed-loop performance of DAD-DPC can vary significantly depending on the specific design and tuning of these components.
To guide practical implementation, we introduce the following heuristic property:

\begin{itemize}
    \item \textbf{Informal Property 2}: 
    Decreasing $\sigma$ leads to a disturbance bound $\mathcal{D}(\sigma)$ that is increasing in size (with respect to set inclusion),
    resulting in a tendency for the system controlled by $\pi(z_t, \mathcal{D}(\sigma))$ to experience fewer constraint violations.
\end{itemize}

This informal property requires only a roughly monotonic relationship between $\mathcal{D}(\sigma)$, $\pi(z_t, \mathcal{D}(\sigma))$, and the resulting violation ratio. It introduces a heuristic violation feedback mechanism that adaptively adjusts the disturbance bound based on recent performance.
Specifically, when the recent violation frequency exceeds the target $\alpha$, $\alpha_t$ decreases in DAD-DPC. According to Property 2, this results in a larger disturbance bound and thus a more conservative control policy, reducing further violations. This adaptive mechanism regulates the average violation rate through feedback. 

In practice, controllers designed to satisfy Informal Property 2 may also lead to empirical satisfaction of \change{the bound}~\eqref{eqn:cons_vio_asy}, even without activating the backup controller \change{$\pi^B$}. 
\change{Notably, the average violations in the closed loop may converge to the bound level $\alpha$, or remain below it, depending on specific scenarios.}
This behavior is supported by some empirical results presented in Sections~\ref{sect:simulation}, ~\ref{sect:experiment} \change{and Appendix~\ref{appen:exp_result}}, which show smooth closed-loop performance and effective regulation of violation levels.
% Note that Theorem~\ref{thm:bound_suf_2} only proves that Property 1 is a sufficient condition for the satisfaction of violation bound~\eqref{eqn:cons_vio_asy}.

% This mechanism forms the basis for defining sufficient conditions to guarantee the satisfaction of~\eqref{eqn:cons_vio_asy}, which will be elaborated in the next section. 

% Any heuristic controller satisfying Informal Property 2 can be adopted, and its specific design largely impacts the closed-loop control performance, providing a basis for engineering-level tuning.

\subsection{A specific design of DAD-DPC for buildings} \label{sect:method_design}
% This section details the application of DAD-DPC for building climate control. The design of its two major components is presented: a robust DPC method based on Willems' Fundamental Lemma, and a disturbance quantification method using conformal prediction.

This section presents a practical implementation of the DAD-DPC components, $\mathcal{D}(\sigma)$ and $\pi(z_t, \mathcal{D}(\sigma))$, tailored for building climate control. This design satisfies Properties 1 and 2 and forms the basis for the experimental validation in Sections~\ref{sect:simulation} and~\ref{sect:experiment}.

The backup controller \change{$\pi^B$} is chosen from default setpoint-based controllers commonly available in commercial HVAC systems. With a conservatively designed rule designed by domain knowledge, it satisfies Property 1. Representative examples of such controllers are detailed in Section~\ref{sect:simulation_default}.

The remainder of this section presents a feasible design of $\mathcal{D}(\sigma)$ and $\pi(z_t, \mathcal{D}(\sigma))$ when $\sigma \neq 0$ satisfying Informal Property 2.
Specifically, we formulate a robust DPC based on Willems' Fundamental Lemma, and describe a disturbance quantification method using conformal prediction.
This results in a data-driven and computationally efficient control scheme.

% (workflow, and complexity)  is discussed in Section 
% which was used in Simulation and experiments
% other data-driven methods (?)

\subsubsection{Data-driven Predictive Control} \label{sect:method_dpc}

A Hankel matrix $H_u$ of depth $L$, constructed from a historical $T$-step input sequence $\{u_t\}_{t=1}^T$, is defined as:
\begin{align*}
    H_u :=
    \begin{bmatrix}
    u_1 & u_2 & \dots & u_{T-L+1}\\
    u_2 & u_3 & \dots & u_{T-L+2}\\
    \vdots & \vdots & &\vdots\\
    u_{L} & u_{L+1} &\dots & u_T
    \end{bmatrix}\;.
\end{align*}
Similarly, Hankel matrices $H_y$ and $H_w$ are constructed for historical outputs $\{y_i\}_{i=1}^T$ and external inputs $\{w_i\}_{i=1}^T$.
An input sequence $\{u_i\}_{i=1}^T$ is called \textit{persistently exciting} of order $L$ if its Hankel matrix $H_u$ has full row rank.
Willems' Fundamental Lemma enables data-driven trajectory prediction without requiring explicit system identification, introduced as follows.

\begin{lemma}\label{lemma:funda}\cite[Theorem 1]{willems2005note}
Consider a controllable linear system, $x_{t+1} = Ax_t + Bu_t$ and $y_t = Cx_t + Du_t$, with $n_x$ as the system's order. If $\{u_i\}_{i=1}^T$ is persistently exciting of order $L + n_x$,
then $\text{colspan}(\begin{bmatrix} H_u^\top & H_y^\top \end{bmatrix}^\top) = \mathfrak{B}_L(A,B,C,D)$, where $\mathfrak{B}_L(A,B,C,D)$ represents  the set of all $L$-step trajectories produced by the linear system.
\end{lemma}

Although this lemma is originally established for noise-free linear systems, extensions to systems with noisy measurements and nonlinear dynamics have been developed~\cite{dorfler2022bridging,lian2023adaptive,yin2024data}.
Furthermore, many extensions have been experimentally validated in various building types for diverse climate control tasks~\cite{chinde2022data,lian2023adaptive,shi2025adaptive,yin2024data}. 

% y_{t-t_{init}+1:t} :
In this paper, we propose a robust bi-level DPC (RB-DPC) scheme leveraging this lemma for the application of DAD-DPC in building climate control.
At time $t$, define $\mathbf{y}_{init} := [y_{t-t_{init}+1}^{\top},\dots,y_{t}^{\top}]^{\top}$ as the previous $t_{init}$-step measurements, $\mathbf{y}_{pred} :=[y_{0|t}^{\top},\dots,y_{N-1|t}^{\top}]^{\top}$ as the $N$-step predicted output sequence, with similar definitions for $u$ and $w$.
The Hankel matrices are constructed with depth $L = t_{init} + N$, separated into components for $t_{init}$ and $N$ steps, such as $H_{y} = [H_{y,init}^\top \,,\,  H_{y,pred}^\top]^\top$ for $y$.
The RB-DPC policy is defined as:
\begingroup
\allowdisplaybreaks
\begin{subequations}\label{eqn:dpc_robust}
\begin{align}
    \pi(z_t,&\mathcal{D}(\sigma)): =u_{0|t}^{\ast}, \text{if }  \sigma \neq 0
    \nonumber \\
     \{\mathbf{u}_{pred}^{\ast},&\mathbf{y}_{pred}^{\ast}\} \, = \argmin_{\mathbf{u}_{pred},\mathbf{y}_{pred}}\;\; \sum_{i=0}^{N-1} I(y_i, u_i) + Q_{\delta}\lVert \delta_i \rVert_2^2 \nonumber \\
        \text{s.t.} 
        &\;F_{y,t+i} \; (y_{i|t} + d_{i}) \leq f_{y,t+i} + \delta_i \; ,  \nonumber \\
        & \quad \quad \forall d_{i} \in \mathcal{D}_i(\sigma) \; , \; \forall i \in \{ 0,\dots,N-1\} \; ,   \label{eqn:dpc_ycons}\\
        &\; u_{i|t} \in \mathcal{U}, \; \forall i \in \{ 0,\dots,N-1\} \; , \label{eqn:dpc_ucons}\\        
        &\;\mathbf{y}_{pred} = H_{y,pred}g \; ,\label{eqn:dpc_pred_y}\\
        &g = \argmin_{g_l,\delta_y} \frac{1}{2}\lVert\delta_y\rVert^2+\frac{1}{2}g_l^\top Q_g g_l\label{eqn:dpc_pred_g}\\
        &\quad\quad\quad\text{s.t.}\;\begin{bmatrix}
        H_{y,init}\\ H_{u,init}\\
        H_{w,init}\\
        H_{u,pred}\\
        H_{w,pred}
        \end{bmatrix}g_l=\begin{bmatrix}
        \mathbf{y}_{init}+\delta_y\\\mathbf{u}_{init}\\ \mathbf{w}_{init}\\\mathbf{u}_{pred}\\\mathbf{w}_{pred}
    \end{bmatrix} \nonumber \;.
\end{align}
\end{subequations}
\endgroup
In this DPC setting, $z_t: = [\mathbf{y}_{init}^{\top}, \mathbf{u}_{init}^{\top}, \mathbf{w}_{init}^{\top}]^{\top}$, and the disturbance bound estimator $\mathcal{D}(\sigma)$ contains $N$ sub-estimators $\mathcal{D}_i(\sigma)$ for each predicted output $y_{i}$. 
\change{\eqref{eqn:dpc_ycons} represents polyhedral inequalities, where $F_{y,t}$ and $ f_{y,t}$ model time-varying comfort constraints caused by occupant movement, as exemplified in~\eqref{eqn:prob_ycons}.}

In the low-level optimization problem, the decision variables are $g_l \in \mathbb{R}^{T-L+1}, \delta_y \in \mathbb{R}^{t_{init}n_y}$, where $n_y$ is the output dimension. 
The regularization cost with the parameter $Q_g$ is used to mitigate the effects of measurement noise and system nonlinearity~\cite{lian2023adaptive}.
Given the previous measurements $\mathbf{u}_{init},\;\mathbf{y}_{init}$ and $\mathbf{w}_{init}$, the equations~\eqref{eqn:dpc_pred_y} and~\eqref{eqn:dpc_pred_g} predict the output sequence $\mathbf{y}_{pred}$ for specified control inputs $\mathbf{u}_{pred}$ and predictive external inputs $\mathbf{w}_{pred}$.

In the high-level optimization problem, the controller optimizes the user-defined control objective subject to input and output constraints. 
This RB-DPC directly applies constraint tightening on the output prediction $\mathbf{y}_{pred}$, as shown in~\eqref{eqn:dpc_ycons} (similar to~\cite{berberich2020robust}).
The extent of constraint tightening is determined by the disturbance bound estimator $\mathcal{D}_i(\sigma)$ and the confidence level $\sigma$. In this paper, we design $\mathcal{D}_i(\sigma)$ using conformal prediction, a method detailed in the next section. This approach enables a data-driven estimation of overall output uncertainties arising from various sources, including model uncertainties and measurement noise. In contrast, previous works such as~\cite{lian2023adaptive,shi2025adaptive} primarily address robustness against weather forecast errors.
In addition, to ensure recursive feasibility, soft output constraints are included, with a quadratic penalty applied using the weight $Q_{\delta}\in\R$. This parameter also allows the controller to adjust violations, particularly in the nominal case $\pi(z_t,\mathcal{D}(1))$.

% Unlike the robust formulation of the bi-level DPC in~\cite{lian2023adaptive,shi2025adaptive},

This bi-level DPC formulation~\eqref{eqn:dpc_robust} is well-suited for building climate control due to several advantages. It is insensitive to the choice of model order and does not require an observer~\cite{yin2024data}. Its linear structure ensures both data and computational efficiency. Compared to standard single-level DPC formulations based on Willems' Fundamental Lemma, the bi-level framework significantly reduces tuning efforts~\cite{shi2025adaptive}. Moreover, it supports efficient adaptive updates, allowing it to approximate varying building operational conditions and enhance representational capability~\cite{shi2023efficient}.
Empirical validations further demonstrate the effectiveness of this bi-level DPC~\cite{lian2023adaptive,shi2025adaptive}. For example, \cite{shi2025adaptive} employed a similar DPC scheme using ten days of operational data for a two-month demand response task, achieving a 29\% cost compared to an industrial rule-based method. 
Therefore, these attributes highlight the bi-level DPC as a practical and efficient solution for building climate control with low commissioning efforts.

% Therefore, it provides a powerful building controller with low commissioning efforts.

% Furthermore, [10] recently proposed using Willems' Fundamental Lemma to build a linear \ac{dpc} directly from \ac{io} data, skipping the modeling and observer steps. It proves a data-efficient solution for building control and is insensitive to the choice of model order [11-12]. 
% For example, [11] constructed a bi-level variant of this \ac{dpc} using two-day operational data, which reduced 18\% energy consumption compared to an industrial rule-based method. Its prediction capability is improved by adaptive data updates to capture time-varying building dynamics, which was validated in a complex two-month demand response experiment [12].

\subsubsection{Disturbance quantification using Conformal Prediction} \label{sect:method_cp}

This section uses conformal prediction to construct $\mathcal{D}_i(\sigma)$ in~\eqref{eqn:dpc_ycons}, which serves as the bound estimator of disturbance $d_{i}$ associated with each predicted output $y_{i}$. The estimator $\mathcal{D}_i(\sigma)$ is parameterized by a specified confidence level $\sigma$ and forms the second key component of the DAD-DPC framework.

Conformal prediction is a distribution-free, data-driven method for generating prediction intervals for regression models~\cite{cp_vovk2005,cp_intro_angelopoulos2021}. In this paper, we employ \ac{scp}~\cite{cp_vovk2005,cp_intro_angelopoulos2021} to construct  $\mathcal{D}_i(\sigma)$.

First, the SCP algorithm is used to estimate $\mathcal{D}_i^{(j)}(\sigma)$, corresponding to the $j$-th dimension of $d_{i}$, i.e. $d_{i}^{(j)}$, for each prediction step $i \in \N_0^{N-1}$. The process is outlined in Algorithm~\ref{alg:cp_w}, where a user-defined controller $\pi^B$ collects the necessary \ac{io} data. For building systems, a rule-based controller is often a practical choice.

\begin{algorithm} 
  \caption{Disturbance Quantification by SCP} \label{alg:cp_w} 
{
\textbf{Input:} Prediction step $i$, dimension $j$, Hankel matrices $H_y, H_u, H_w$, calibration size $n_{cal}$, controller $\pi^B$ for data collection\\
\textbf{Output:} Function $\mathcal{D}_i^{(j)}(\sigma)$ for $(1-\sigma)$-confidence bound of $d_{i}^{(j)}$, where $\sigma \in [0,1)$
\begin{itemize} [nolistsep,leftmargin=1.5em]
    \item[1)] Use $\pi^B$ to control the target system starting at time $t_{c}$ and collect \ac{io} data of length $T_c:=n_{cal}+n_{init}+N-2$.
    \item[2)] For each time $t \in \N_{t_c+t_{init}-1}^{t_c + T_c - N}$,
    compute the output prediction $\mathbf{y}_{pred}$ by~\eqref{eqn:dpc_pred_y} and~\eqref{eqn:dpc_pred_g}, based on actual control and external inputs, i.e. $\mathbf{u}_{pred} = [u_{t},\dots,u_{t+N-2}]$,$\mathbf{w}_{pred} = [w_{t},\dots,w_{t+N-2}]$.
    In addition, compute the prediction residual: 
    % {\setlength{\abovedisplayskip}{1.6pt}\setlength{\belowdisplayskip}{1.6pt}
    \begin{align} \label{eqn_cp_residual}
        R_{d_{i}^{(j)}}(t) = &\lvert y_{t+1+i}^{(j)} - y_{i|t}^{(j)}\rvert \; .
    \end{align}
    % }    
    \item[3)] Construct the function $\mathcal{D}_i^{(j)}(\sigma) $:
    \begin{equation} \label{eqn:cp_bound_1}
        \begin{aligned}
        \mathcal{D}_i^{(j)}(\sigma)  = 
        [-q&_{d_{i}^{(j)}}(\sigma), +q_{d_{i}^{(j)}}(\sigma)] \; ,
        \end{aligned}
    \end{equation}     
    \noindent where $q_{d_{i}^{(j)}}(\sigma)$ is the $\lceil n_{cal}(1-\sigma)\rceil$-th smallest residual in $\left\{R_{d_{i}^{(j)}}(t_c+t_{init}-1), \dots, R_{d_{i}^{(j)}}(t_c + T_c - N) \right\}$.
\end{itemize}
}
\end{algorithm}

Second, the estimator $\mathcal{D}_i(\sigma)$ is constructed by aggregating the individual $\mathcal{D}_i^{(j)}(\sigma)$ for all dimensions:
\begin{equation} \label{eqn:cp_bound_tot}
\begin{aligned}
\mathcal{D}_i&(\sigma)  = \\ & \begin{cases} 
        \left\{d \middle| 
\begin{aligned}
    & d^{(j)}  \in \mathcal{D}_i^{(j)}(\sigma) \; , \\ &\text{for} \;  j \in \{1,\dots,n_y \} 
\end{aligned}\right\}, & \text{if } 1-\sigma \in (0,1] \; ; \\
        \{\mathbf{0}\} \; , & \text{if } 1-\sigma = 0 \; .     
    \end{cases} 
\end{aligned}    
\end{equation}
Therefore, for $1-\sigma >0$, the estimated disturbance bound $\mathcal{D}_i(\sigma)$ is a box constraint derived from the prediction residuals in~\eqref{eqn_cp_residual}. This approach enables the estimation of an overall output uncertainty bound that accounts for multiple uncertainty sources, including model inaccuracies and measurement noise.

\subsubsection{The resulting DAD-DPC}\label{sect:result_daddpc}

% online, offline process description
We now summarize the complete DAD-DPC design tailored for building climate control. As illustrated in Figure~\ref{fig:dad_concept}, it comprises offline and online stages. In the offline stage, past \ac{io} data are used to construct the RB-DPC~\eqref{eqn:dpc_robust} and the disturbance bound estimator $\mathcal{D}_i(\sigma)$~\eqref{eqn:cp_bound_tot}. During online operation, the system is controlled according to Algorithm~\ref{alg:dad_dpc}, using these components and along with the default controller \change{$\pi^B$}.

By construction, the use of default HVAC controllers for \change{$\pi^B$} ensures satisfaction of Property 1, which in turn provides a sufficient guarantee for the violation bound~\eqref{eqn:cons_vio_asy}, as established in Theorem~\ref{thm:bound_suf_2}. As for Informal Property 2, $\mathcal{D}_i^{(j)}(\sigma)$ in~\eqref{eqn:cp_bound_1} increases as $\sigma$ becomes larger, implying that $\mathcal{D}(\sigma)$ also expands. This leads to tighter output constraints in RB-DPC, which in turn tends to reduce the frequency of violations. Notably, when applying this DAD-DPC controller, in several simulation and experimental cases presented in Sections~\ref{sect:simulation}, \ref{sect:experiment} the controller design based on Property 2 alone was sufficient to empirically satisfy the violation bound~\eqref{eqn:cons_vio_asy} without activating the backup controller \change{$\pi^B$}.

\begin{remark}[Computational complexity] \label{remark:computation}
The disturbance bound estimator~\eqref{eqn:cp_bound_tot} derived from conformal prediction introduces box constraints on $d_{i}$ in the RB-DPC~\eqref{eqn:dpc_robust}. When polyhedral output constraints,  such as the comfort constraint~\eqref{eqn:prob_ycons} commonly used in buildings climate control, are applied, the computational complexity of $\pi(z_t,\mathcal{D}(\sigma))$ remains equivalent to that of the nominal controller $\pi(z_t,\mathcal{D}(1))$. \change{If, in addition, the input constraint set $\mathcal{U}$ is polyhedral and} $I(y_i, u_i)$ is a positive-definite quadratic function or an affine function, \eqref{eqn:dpc_robust} is quadratic programming \change{(QP), which can be efficiently solved using QP solvers such as Gurobi~\cite{gurobi} and MATLAB Quadprog~\cite{matlab:quadprog}.}
\end{remark}

\begin{remark} [Other data-driven methods]
Theorem~\ref{thm:bound_suf_2} does not depend on specific choices of $\pi(z_t,\mathcal{D}(\sigma))$ and $\mathcal{D}(\sigma)$. This flexibility allows the construction of DAD-DPC using alternative predictive models and disturbance quantification techniques. For example, the predictor in the RB-DPC~\eqref{eqn:dpc_robust} could be replaced with other data-driven models, such as a Gaussian Process (GP) model and neural networks. The disturbance bound estimator $\mathcal{D}(\sigma)$ could be derived from the GP-based uncertainty quantification or scenario-based methods. Exploring these alternatives is a promising direction, particularly for building systems with complex nonlinear dynamics.
\end{remark}

% It is worth noticing that the DAD-DPC framework can be extended to other DPC methods, such as the NNs. Although a more complex data-driven model can lead to nonconvex robust optimization problems, some techniques, such as Real-time iterations and Practical MPC method, provide efficient suboptimal solutions. Therefore, it is interesting for future exploration.

\section{Case studies: simulation} \label{sect:simulation}
This section presents three simulation case studies conducted to validate the efficacy of the proposed DAD-DPC framework. The simulation setups are briefly described in Section~\ref{sect:simu_setup}, followed by the default controller setup and the parameter choices for DAD-DPC. Finally, the results are presented and discussed in Section~\ref{sect:simu_result}

% \vspace{-2mm}
\subsection{Setup of simulation cases}
\label{sect:simu_setup} % Message: different setups

The simulation study leverages the high-fidelity Building Optimization Testing Framework (BOPTEST)~\cite{blum2021building}, which provides diverse cases for benchmarking building control performance. Three cases with distinct configurations were deployed to validate the DAD-DPC method. The setups of these cases, including building architecture, HVAC systems, occupancy schedules, and climate data, are summarized \change{in Table~\ref{tab:case_setup}}.
\change{Further details for these emulators are publicly available. 
Case~1 (``BESTEST hydronic heat pump'') is provided directly by BOPTEST\footnote{\url{https://ibpsa.github.io/project1-boptest/testcases/index.html}}. 
Case~2 (``single zone commercial hydronic'') and Case~3 (``Multi-zone office floor'') were provided through the BOPTEST Challenge~\cite{boptestChallenge}\footnote{\url{https://adrenalin.energy/BOPTEST-Challenge-Smart-building-HVAC-control}}. 
Specifically, Case~2 is a modified version of the original ``single zone commercial hydronic,'' and Case~3 is a modified version of the original ``multizone office simple air''\footnotemark[1]. 
Both modified FMUs and their descriptions are available online\footnote{\url{https://data.sintef.no/feature/fe-7d9861ec-ae6e-4947-ae43-85f27fee3cd3}} and can be imported into the BOPTEST package as described in the Challenge instructions.}

\begin{table*}[!ht]
{\color{black}
\centering
\caption{Setup for Three Cases}
\label{tab:case_setup}
\renewcommand{\arraystretch}{0.2}
\begin{tabularx}{\textwidth}{@{}lYYY@{}}
\toprule
\textbf{Specification } & \textbf{Case 1: BESTEST Hydronic Heat Pump} & \textbf{Case 2: Single Zone Commercial Hydronic} & \textbf{Case 3: Multi-Zone Office Floor} \\ \midrule
\textbf{Architecture} &
Single-zone residential dwelling (family of five) based on BESTEST case 900; rectangular plan $12\times16$\,m; height 2.7\,m. &
Real building abstraction ($8500\,\mathrm{m}^2$) with classrooms (40\%), study (25\%), offices (15\%), common spaces (20\%); basement houses HVAC; simulated as a single zone. &
Middle floor of an office building in Chicago; five zones (four perimeter + one core). See Fig.~\ref{fig:case3_5zone}. \\ \midrule
\addlinespace[0.2em]
\textbf{HVAC system} &
Air-to-water modulating heat pump serving hydronic floor heating. &
District heating supply with main circulation pump; one AHU and one radiator; two control valves regulate flow to AHU coil and radiator circuits; rotary heat-recovery wheel. &
Similar to the one used in Case~2 (Difference:
five radiators and valves for five rooms).\\ \midrule
\addlinespace[0.2em]
\textbf{Nominal capacities} & Heat pump: 15\,kW. & District heating: 500\,kW; AHU coil: 250\,kW;
radiator: 250\,kW. & District heating: 90\,kW; AHU coil: 37\,kW; radiator: 50\,kW. \\ \midrule
\addlinespace[0.2em]
\textbf{Occupancy} &
Five occupants; occupied time for control: before 07{:}00 and after 20{:}00 on weekdays, continuous on weekends. &
Occupancy modeled from camera-based historical data from the
real building; occupied time for control: 07{:}00–19{:}00 on weekdays. &
Fixed weekday/weekend pattern with Gaussian noise; occupied time for control: 07{:}00–19{:}00 on weekdays. \\ \midrule
\addlinespace[0.2em]
\textbf{Climate data} &
One year of weather for Brussels, Belgium. &
One year of weather for Copenhagen, Denmark. &
One year of weather for Chicago, USA. \\
\bottomrule
\end{tabularx}
}
\end{table*}

\begin{table*}[!ht]
\centering
\caption{Summary of Main Differences among the Three Cases}
\label{tab:simulation_cases}
\renewcommand{\arraystretch}{0.5}
\begin{tabularx}{0.75\textwidth}{@{}M C{20MM} C{20MM} C{41MM} C{22MM}@{}}
\toprule
\textbf{Specification} & \textbf{Building Type}      & \textbf{Total Area ($m^2$)} & \textbf{HVAC Systems} & \textbf{Number of Zones} \\ \midrule
\textbf{Case 1}        & Residential                 & 192                         & Heat pump \& floor heating & 1                        \\ \midrule
\textbf{Case 2}        & School                      & 8500                        & District heating \& AHU \& radiator & 1                        \\ \midrule
\textbf{Case 3}        & Office                      & 1662.5                      & District heating \& AHU \& radiator & 5                        \\ \bottomrule
\end{tabularx}
\end{table*}

% \begin{table*}[!ht]
% \centering
% \caption{Summary of Main Differences among the Three Cases}
% % \label{tab:simulation_cases}
% \begin{tabular}{|c|c|c|c|c|}
% \hline
% \textbf{Specification} & \textbf{Building Type}      & \textbf{Total Area ($m^2$)} & \textbf{HVAC Systems} & \textbf{Number of Zones} \\ \hline
% \textbf{Case 1}        & Residential                 & 192                         & Heat pump \& floor heating & 1                        \\ \hline
% \textbf{Case 2}        & School                      & 8500                        & District heating \& AHU \& radiator & 1                        \\ \hline
% \textbf{Case 3}        & Office                      & 1662.5                      & District heating \& AHU \& radiator & 5                        \\ \hline
% \end{tabular}
% \end{table*}

% Explain the difference from the table. Highlight that we want to test on different types. Case 2 and 3 are from the training stage of BOPTEST challenges (why? multizone?)
The objective of this simulation study is to empirically validate the DAD-DPC framework across three distinct simulation cases. Key differences in the specifications of these cases are summarized in Table~\ref{tab:simulation_cases}. Differences in building types and total areas result in varying occupant numbers and schedules.
In terms of HVAC systems, Cases 2 and 3, derived from the BOPTEST challenge, employ multi-input systems, whereas Case 1 features a single-input configuration.
Furthermore, unlike Cases 1 and 2, Case 3 models the building with five distinct zones, making it a multi-output system.

% Explain the weather forecast and occupancy, the measurement, comfort (later in DPC).
BOPTEST offers several APIs to facilitate simulation. The forecast API predicts weather conditions, occupant presence, and internal heat gains based on the current simulated time. In this study, only weather forecasts were utilized, as occupant and heat gain forecasts are less commonly available in real-world applications. Additionally, the results API retrieves \ac{io} measurements, while the KPI API calculates cumulative HVAC energy costs.

\begin{figure}[!ht]
    \centering
    \includegraphics[width=0.6\linewidth]{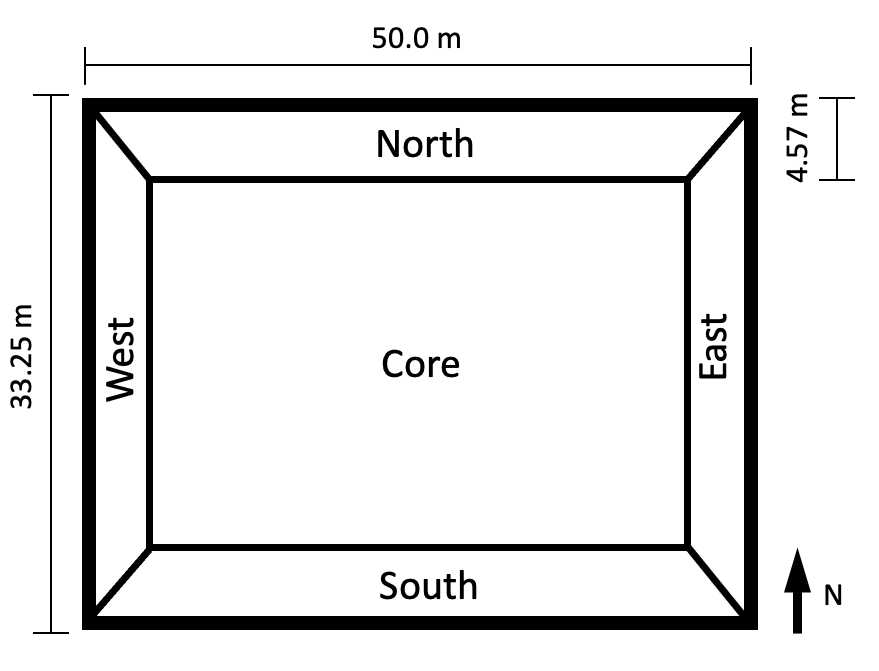}
        \caption{Illustration of five zones in Case 3~\cite{deru2008doe}.}
    \label{fig:case3_5zone}
    % \vspace{1em}
\end{figure}

% \vspace{-2mm}
\subsection{Setup of controllers} 

The same type of comfort constraint, $\mathcal{Y}_t$, is applied across all three cases and is defined as follows:
\begin{align} 
    \mathcal{Y}_t= \begin{cases}
       \{y| y 
 \geq {21}I_{n_y} {^\circ} C \} \; , &  \text{during occupied periods} \; ; \\
        \{y| y 
 \geq {18}I_{n_y} {^\circ} C \} \; , & \text{otherwise} \; ,
    \end{cases}
\end{align}
where the occupancy schedules for each case were detailed in the previous section. In the following, we detail the controller settings used for building climate control in the simulation.

\begin{table*}[!ht]
{\color{black}
\centering
\caption{Default controllers in the three cases}
\label{tab:default_ctrl_summary}
\renewcommand{\arraystretch}{0.2}
\setlength{\tabcolsep}{6pt}
\begin{tabularx}{\textwidth}{@{}l Z Y Y Y@{}}
\toprule
\textbf{Case} & \textbf{PI loops} & \textbf{Controlled variables and setpoints} & \textbf{Manipulated inputs} & \textbf{Preheat policy} \\
\midrule
\textbf{Case 1} &
1 &
Zone temperature (occupied: $21.2^\circ$C; unoccupied: $20.5^\circ$C) &
HP modulation signal; evaporator fan and floor heating system pump interlocked with HP on/off &
-- \\ \midrule
\addlinespace[0.2em]
\textbf{Case 2} &
3 &
CO$_2$ $<950$\,ppm; AHU supply air $21^\circ$C; zone temperature (occupied: $22^\circ$C; unoccupied: $18^\circ$C) &
AHU supply fan speed; AHU coil valve; radiator valve &
Raise to $22^\circ$C starting 2\,h before occupancy (4\,h on Mondays) \\ \midrule
\addlinespace[0.2em]
\textbf{Case 3} &
7 
&
Same as Case~2, except five zone-temperature setpoints (all $22^\circ$C), one per zone on the multi-zone floor &
Same as Case~2, except five radiator valves (one per zone) &
-- \\
\bottomrule
\end{tabularx}
}
\end{table*}

\subsubsection{Default controller} \label{sect:simulation_default}
The default controllers provided by the three BOPTEST cases are described \change{in Table~\ref{tab:default_ctrl_summary}}. These controllers serve as baselines for comparison with the proposed DAD-DPC controllers.
\change{Cases~2 and~3 employ three classes of PI controllers: (i) indoor CO$_2$ maintained below $950$\,ppm via AHU supply fan speed, (ii) AHU supply air temperature regulated to $21^\circ$C via the AHU valve, and (iii) zone temperature regulated via radiator valve(s).
In Case~3, the third class is designed for each zone (five PI controllers, one per radiator valve).
Across all three cases, the occupied temperature setpoints are slightly above the comfort bound ($21^\circ$C) to mitigate oscillation-induced discomfort (Case~1: $21.2^\circ$C; Cases~2--3: $22^\circ$C).}

% \textbf{Case 1:}  A local PI controller regulates the HP's modulating signal based on feedback from the indoor temperature sensor. The evaporator fan and the floor heating system pump are activated only when the HP is on.
% The default controller sets the temperature to $21.2^\circ C$  when occupied and $20.5^\circ C$  when unoccupied. These setpoints are chosen to reduce discomfort caused by oscillations around the target temperature.

% \textbf{Case 2:} This case employs three PI controllers. The first controller maintains indoor CO2 concentration below $950$ ppm by adjusting the AHU supply fan speed. The second controller regulates the AHU supply air temperature to $21^\circ C$ by controlling the AHU valve position. The third controller controls the indoor temperature by adjusting the radiator valve position.
% The default indoor temperature setpoint is $18^\circ C$ during unoccupied periods. Two hours before occupancy (four hours on Mondays), the setpoint is raised to $22^\circ C$ to reduce comfort violations.

% \textbf{Case 3:} This case uses three PI controllers with configurations and setpoints similar to those in Case 2.

In the DAD-DPC implementation, each case used its own default control policy as  \change{$\pi^B$}, with Cases 2 and 3 using their original configurations and Case 1 employing the default controller with a $22^\circ$C setpoint. \change{The sets $\mathcal{Y}_{\text{lim}}$ were all chosen as a suitable operating range, $\{y|{15}^\circ C \leq y \leq {30}^\circ C \}$.}

\subsubsection{Setup of DAD-DPC}

% (1) table: y (Case 1, 2: indoor temperature of one zone; Case 3: indoor temperatures of five zones), u (Case 1: HP electrical power; Case 2, 3: AHU heating power, radiator heating power), w (Case 1,2,3: outdoor temperature, radiation)
% Building (y), HVAC (u). Describe the input and output measurement,

In this paper, DAD-DPC was implemented for temperature control, while the ventilation systems continued to operate using the default controller. The \ac{io} configurations for DAD-DPC across the three cases are summarized in Table~\ref{tab:simu_dpc_io}.

\begin{table}[!ht]
\centering
\caption{Summary of \ac{io} for the DAD-DPC of three Cases}
\label{tab:simu_dpc_io}
\renewcommand{\arraystretch}{0.3}
\begin{tabularx}{0.95\columnwidth}{C{10mm} p{6.5cm}}
\toprule
\textbf{Variable} & \textbf{Cases} \\ \midrule
y & Cases 1, 2: indoor temperature of one zone \newline Case 3: indoor temperatures of five zones \\ \midrule
u & Case 1: HP electrical power \newline Cases 2, 3: AHU heating power, radiator heating power \\ \midrule
w & Cases 1, 2, 3: outdoor temperature, radiation \\ \bottomrule
\end{tabularx}
\end{table}

% \begin{table}[!ht]
% \centering
% \caption{Summary of \ac{io} for the DAD-DPC of three Cases}
% % \label{tab:simu_dpc_io}
% \begin{tabular}{|c|p{6.5cm}|}
% \hline
% \textbf{Variable} & \textbf{Cases} \\ \hline
% y & Cases 1, 2: indoor temperature of one zone \newline Case 3: indoor temperatures of five zones \\ \hline
% u & Case 1: HP electrical power \newline Cases 2, 3: AHU heating power, radiator heating power \\ \hline
% w & Cases 1, 2, 3: outdoor temperature, radiation \\ \hline
% \end{tabular}
% \end{table}

For the output $y$, Case 3 includes five dimensions corresponding to the five zones. The external input $w$ comprises the ambient dry-bulb temperature at ground level and global horizontal irradiation, as provided by the BOPTEST forecast and results APIs.
For the control input $u$, heating power was selected for Cases 2 and 3, as the building thermodynamics can be reasonably approximated using linear models, making them suitable for RB-DPC. In Case 1, although nonlinearity arises from the HP's coefficient of performance, RB-DPC effectively approximates short-term behaviors and benefits from adaptive updates for long-term operations~\cite{shi2025adaptive}. 

% Low-level controllers
Notably, the chosen inputs are not directly mapped to the control variables in the BOPTEST cases. For each optimal input computed by DAD-DPC, low-level controllers were employed to translate it into BOPTEST-compatible variables. 
In Case 1, the HP was toggled on and off within each sampling period, with on/off durations adjusted to match the desired average electrical power. Similarly, in Cases 2 and 3, HVAC systems were controlled by modulating setpoints or valve positions to achieve the required heating power.

Across all three cases, the following common parameters were used for DAD-DPC: sampling time, 15 minutes;  prediction horizon $N$, 96 steps; initial steps $t_{init}$, 12 steps; regularization weight $Q_g$, 0.01$I$; regularization weight $Q_{\delta}$, 10; Hankel matrix data length $T$, 672 steps (one week); SCP data length $T_c$, 672 steps (one week); $\mathcal{Y}_{\text{lim}}, \{y|{15}^\circ C \leq y \}$; $\alpha_{0}$, 0; update rate $\eta$, 0.5.  
\change{The sampling time and the prediction horizon were selected based on our previous control experience in the BOPTEST challenge. They are also consistent with the practical MPC guideline for buildings from~\cite{drgovna2020all}: it recommends $10$–$20$ samples over the rise time and reports typical ranges of $15$–$180$ minutes in the literature; the prediction horizon $N$ should be chosen so that $N$ sampling times span the system’s settling time while balancing model and forecast uncertainty.}
The parameters, $t_{init}, Q_g, T$ were chosen based on empirical guidelines from~\cite{shi2025adaptive}. A segmented trajectory trick from~\cite{o2022data} was applied in DPC to reduce data requirements.  
Two weeks of data for $T$ and $T_c$ were collected using a bang-bang controller for each case. A small $\alpha_{0}$ was selected to limit frequent violations at the beginning, while a moderate update rate $\eta$ was chosen to prevent extreme adaptive adjustments.  
Apart from those choices, in the RB-DPC, the control objective was set to $I(y_t,u_t)= u_t$, with input constraints $\mathcal{U}$ defined as the most recent electrical or heating power levels when the respective HVAC system was on.
\change{Under these settings, the resulting RB-DPC controllers were QP problems, as analyzed in Remark~\ref{remark:computation}, and were all solved by Gurobi~\cite{gurobi} in all simulation cases.}
In this simulation study, various $\alpha$ values were tested to validate the satisfaction of the violation bound~\eqref{eqn:cons_vio_asy} using DAD-DPC. The results of this validation are presented in Section~\ref{sect:simu_result}.

% (2) table: time step, DAD-DPC parameters 
% \begin{table}[!ht]
% % \centering 
% \footnotesize
% \begin{center}
% \caption{Choice of parameters for the RB-DPC in three simulation cases} \label{tab:para}
% \begin{tabular}{  l  l}
% % \begin{tabular}{  b{5.5cm}  b{3cm} } %\begin{tabular}{ | b{5cm} | b{1cm}| }
%   \hline
%   No. of Prediction Steps $N_{}$ & 96 \\ 
%   No. of Initial Steps $t_{init}$ & 12\\ 
%   Length of data for Hankel matrices $T$ & 672\\ 
%   Regularization weight $Q_g$ & 0.01$I$ \\  
%   \hline  
% \end{tabular}
%  \end{center}
% \end{table}

\subsection{Results of simulation cases}
\label{sect:simu_result}
To validate the ability of DAD-DPC to satisfy the average comfort violation constraint~\eqref{eqn:cons_vio_asy},  we evaluated its performance over two-week operation periods (1344 steps) and observed its asymptotic behavior. These two-week periods correspond to peak heating days, centered around the day with the highest 15-minute heating load of the year for each case, as determined by the BOPTEST API. The selected dates were: Case 1, January $17^{\text{th}}$ to January $30^{\text{th}}$; Case 2, February $7^{\text{th}}$ to February $20^{\text{th}}$; Case 3, December $15^{\text{th}}$ to December $28^{\text{th}}$. To reflect sensor noise, Gaussian noise $\mathcal{N}(0,0.1)$ was added to the output measurements.

% \subsubsection{Control with different targeted  violation levels}

% Analysis: (1) violation, convergence (2) trade off, 5\%, what is the savings;  (3) bound; 
% (1) Fig: details. $100\%*(1-\bar{\alpha}_t)$
% (2) Fig: energy, violation
% (3) Fig: upper bound, max alpha t

% In three cases, we chose a same updating rate $\eta=0.5$.

\begin{figure}[!ht]
    \centering
    \includegraphics[width=1.0\linewidth]{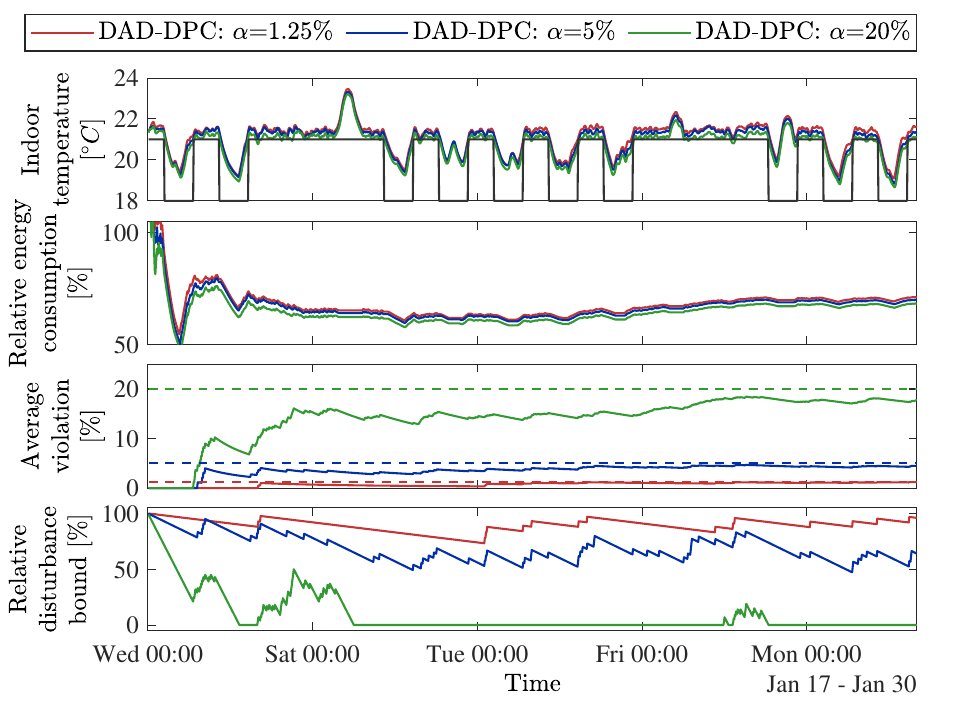}
        \caption{Case 1: trajectories of indoor temperature, relative energy consumption, average violations, relative disturbance bound (sub-figures from top to bottom).}
    \label{fig:case1_trajectory}
    \vspace{1em}
\end{figure}

\begin{figure}[!ht]
    \centering
    \includegraphics[width=1.0\linewidth]{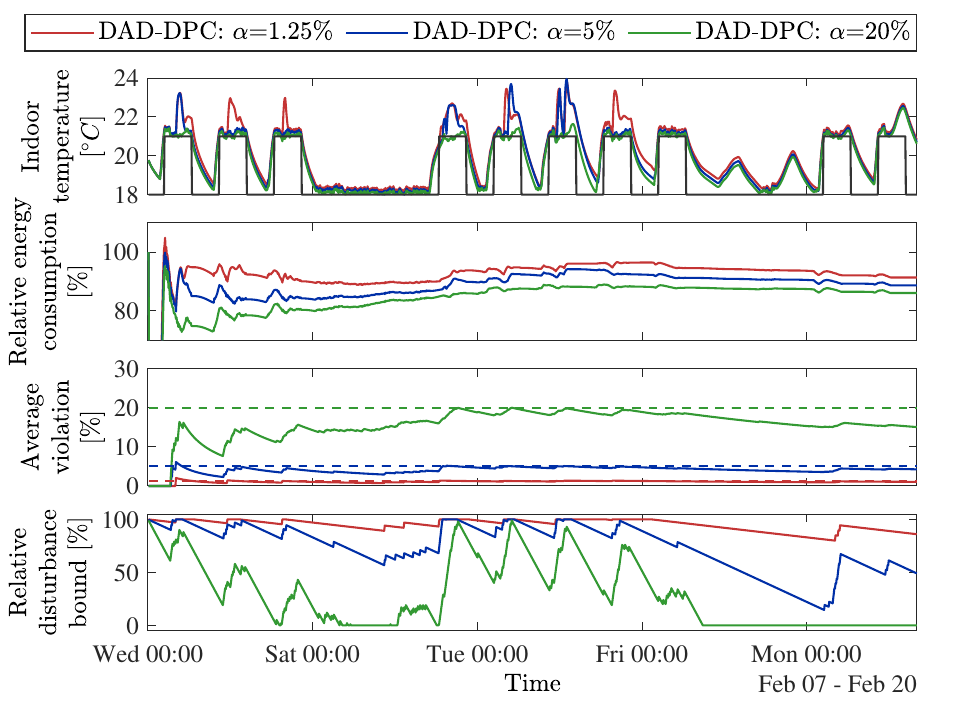}
        \caption{Case 2: trajectories of indoor temperature, relative energy consumption, average violations, relative disturbance bound (sub-figures from top to bottom).}
    \label{fig:case2_trajectory}
    % \vspace{1em}
\end{figure}

\begin{figure}[!ht]
    \centering
    \includegraphics[width=1.0\linewidth]{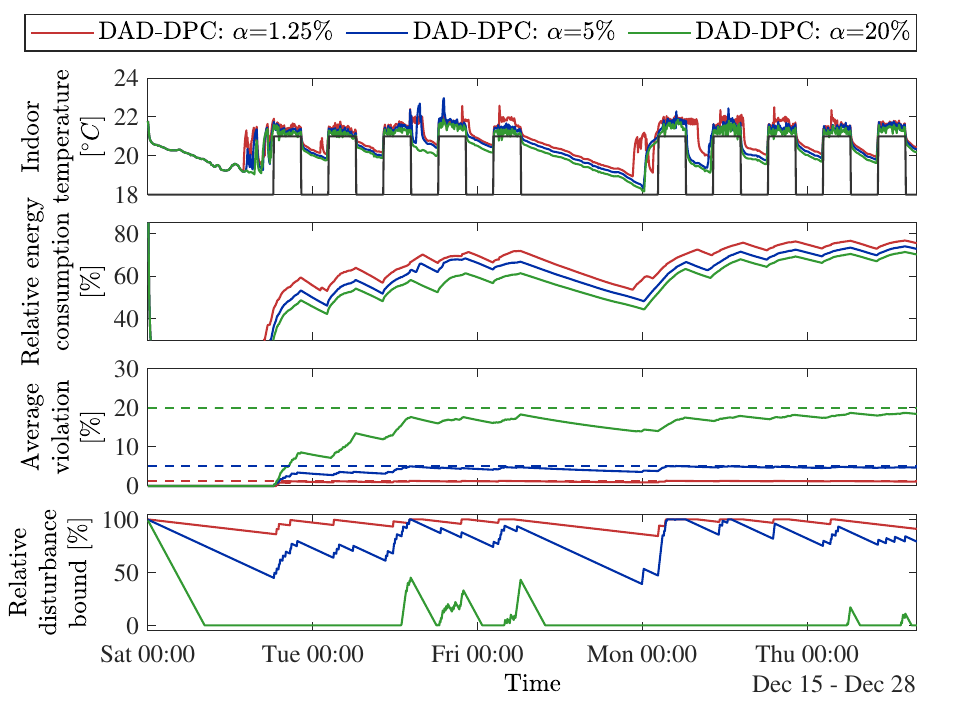}
        \caption{Case 3: trajectories of indoor temperature of the \textbf{west} zone, relative energy consumption, average violations, relative disturbance bound (sub-figures from top to bottom).}
    \label{fig:case3_trajectory}
    \vspace{1em}
\end{figure}

Three values of $\alpha$ ($1.25\%, 5\%, 20\%$) were tested to assess the DAD-DPC framework. Figures~\ref{fig:case1_trajectory}, \ref{fig:case2_trajectory} and~\ref{fig:case3_trajectory} illustrate the two-week trajectories for the three cases. 

The top two sub-figures display the indoor temperature and the relative energy consumption trajectories, calculated as $\frac{\sum_{i=t}^{0} u_i}{\sum_{i=t}^{0} u_i^{\text{default}}}\times 100\%$, where $u_t^{\text{default}}$ is the power consumption of the corresponding default controller.
For most of the time during the three cases, the indoor temperature was effectively maintained near the required comfort constraints. During unoccupied periods, the DAD-DPC relaxed the temperature to save energy. It also preheated zones before occupied periods to ensure comfort.
The total energy consumption across the simulations was consistently lower than that of the default controllers, demonstrating DAD-DPC's capability to achieve energy-efficient building operation.
Moreover, the \ac{io} trajectories exhibited consistent and reasonable behavior across the different $\alpha$ values. Larger $\alpha$ values permitted more comfort violations, leading to indoor temperatures closer to the constraint boundaries and lower energy consumption. At the start of the simulations, the small initial value $\alpha_0=0$ induced conservative behavior, which temporarily resulted in higher energy consumption compared to the default controllers (e.g., Case 1 in Figure~\ref{fig:case1_trajectory}). However, as $\alpha_t$ was adapted, this conservative behavior diminished.

In a few instances (e.g. Case 2 with $\alpha=1.25\%$ and $5\%$, \change{Case 3 with $\alpha=1.25\%$}), the temperature was elevated due to the activation of the conservative backup controller \change{$\pi^B$}.
\change{According to the rule in (4) of Algorithm 1, $\pi^B$ is activated when $y_t \notin \mathcal{Y}_{\text{lim}}$ or $\bar{\alpha}_t=0$.  
In all the simulations, the indoor temperature stayed within the relatively large set $\mathcal{Y}_{\text{lim}}$.
Therefore, the activation of $\pi^B$ in these cases was solely due to $\bar{\alpha}_t=0$, as shown in the bottom sub-figures, which is discussed subsequently.}

The bottom two sub-figures in Figures~\ref{fig:case1_trajectory}, \ref{fig:case2_trajectory} and~\ref{fig:case3_trajectory} illustrate the average violations ($\frac{\sum_{i=0}^{t} v_i}{t} \times 100\%$), and the relative disturbance bound ($(1-\bar{\alpha}_t)\times 100\%$). 
With DAD-DPC, the average violation asymptotically \change{satisfied} the specified level \change{bound} $\alpha$ across all three cases. 
\change{The backup controller $\pi^B$} was applied occasionally \change{when $\bar{\alpha}_t = 0$. In the sub-figures, it corresponds to periods with a $100\%$ relative disturbance bound, such as during the daytime of the second week in Case 2 (Figure~\ref{fig:case2_trajectory}, $\alpha=1.25\%$ and $5\%$). During these periods, $\pi^B$ conservatively increased the indoor comfort away from the lower temperature bound, reducing the average violation in accordance with Property 1.  Once  $\bar{\alpha}_t>0$ again, the input policy switched back to the DPC controller $\pi(z_t,  \mathcal{D}(\bar{\alpha}_t))$, as defined in (4). As a result, $\pi^B$ guaranteed the lower bound of $\alpha_t$, thereby ensuring satisfaction of the average violation bound~\eqref{eqn:cons_vio_asy}, as stated in Theorem~\ref{thm:bound_suf_2}.} 
\change{The DPC controller with adaptive disturbance updates bound through $\bar{\alpha}_t$ also played an important role.} When the violation frequency was below the target $\alpha$, the relative disturbance bound decreased, encouraging more frequent violations due to Property 2, and vice versa. Across some $\alpha$ scenarios, (e.g. Case 1, $\alpha=5\%, 20\%$), \change{$\pi^B$} was not activated except for the initial $\alpha_0$.
Among these scenarios, Property 2 itself ensured a lower bound on \change{$\alpha_{t}$}, thereby satisfying \change{the bound}~\eqref{eqn:cons_vio_asy}. This result also confirmed that Property 1 is a sufficient but not necessary condition in Theorem~\ref{thm:bound_suf_2}. Additionally, based on Lemma~\ref{thm:bound_suf_1}, the strict violation bound~\eqref{eqn:cons_vio_rob} was occasionally satisfied due to the conservative initial setting of $\alpha_0=0$ was chosen (e.g., Case 1, $\alpha=5\%, 20\%$).

% As a result, the higher violation bound $\alpha$ tends to lead to a lower relative disturbance bound, Therefore, the corresponding less conservative operation of the indoor temperature requires less power consumption.

Moreover, the average violation in most cases converged \change{toward} the target bound $\alpha$.
\change{This convergence aligns with thermodynamic principles: operating at lower temperatures with occasional comfort violations generally leads to energy savings.}
This behavior is also consistent with the asymptotic lower bound~\eqref{eqn:cons_vio_asy_lower} provided by Lemma~\ref{thm:bound_suf_1}. Intuitively, this lower bound requires that when $\bar{\alpha}_t$ is large and a small constraint tightening is used in DAD-DPC, the controller results in average violations greater than $\alpha$.

% The non-convergent violation behavior, observed in Case 2, $\alpha = 20\%$ (the green lines in Figure~\ref{fig:case2_trajectory}) can be explained by this mechanism. 
% During the final days of the simulation, warmer temperatures reduced the heating demand, and even the nominal control policy, $\pi(z_t, \mathcal{D}(1))$, failed to induce sufficient violations to meet the target. In such scenarios, choosing a smaller $Q_{\delta}$ could be a practical adjustment to increase violations.

\begin{figure}[!ht]
    \centering
    \includegraphics[width=1.0\linewidth]{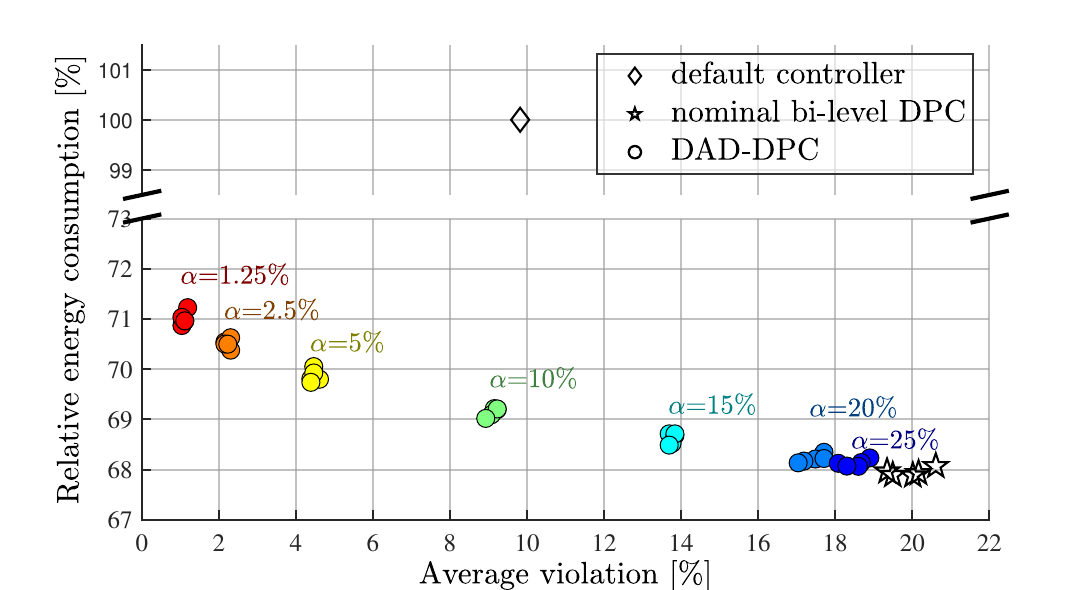}
        \caption{Case 1 during peak heating days: the relative energy consumption and average comfort violations of different controllers. For DAD-DPC, each color corresponds to a different $\alpha$}
    \label{fig:case1_scatter}
    % \vspace{1em}
\end{figure}

\begin{figure}[!ht]
    \centering
    \includegraphics[width=1.0\linewidth]{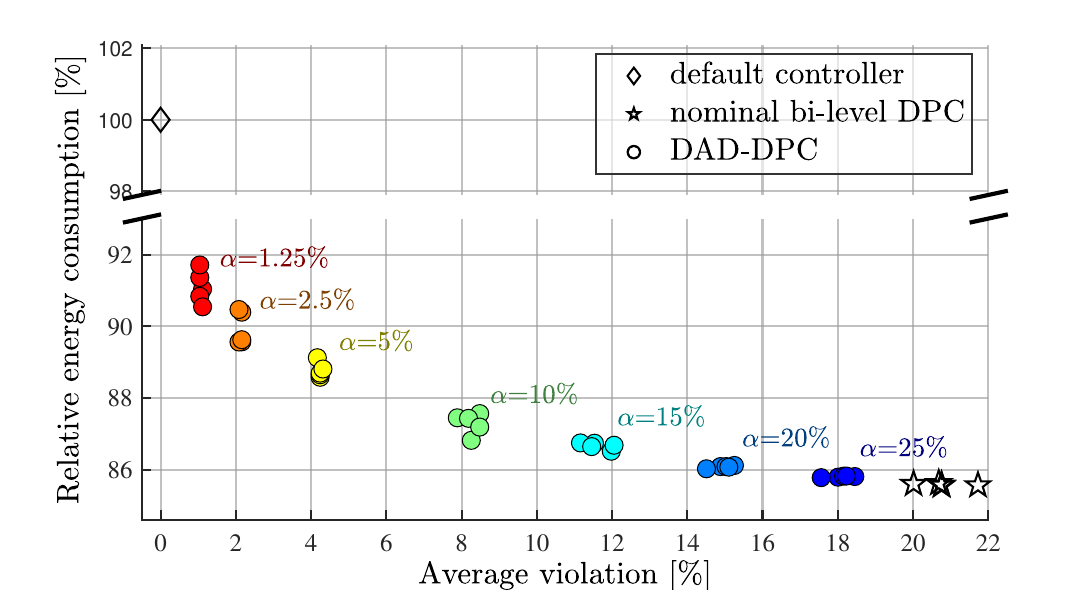}
        \caption{Case 2 during peak heating days: the relative energy consumption and average comfort violations of different controllers. For DAD-DPC, each color corresponds to a different $\alpha$}
    \label{fig:case2_scatter}
\end{figure}

\begin{figure}[!ht]
    \centering
    \includegraphics[width=1.0\linewidth]{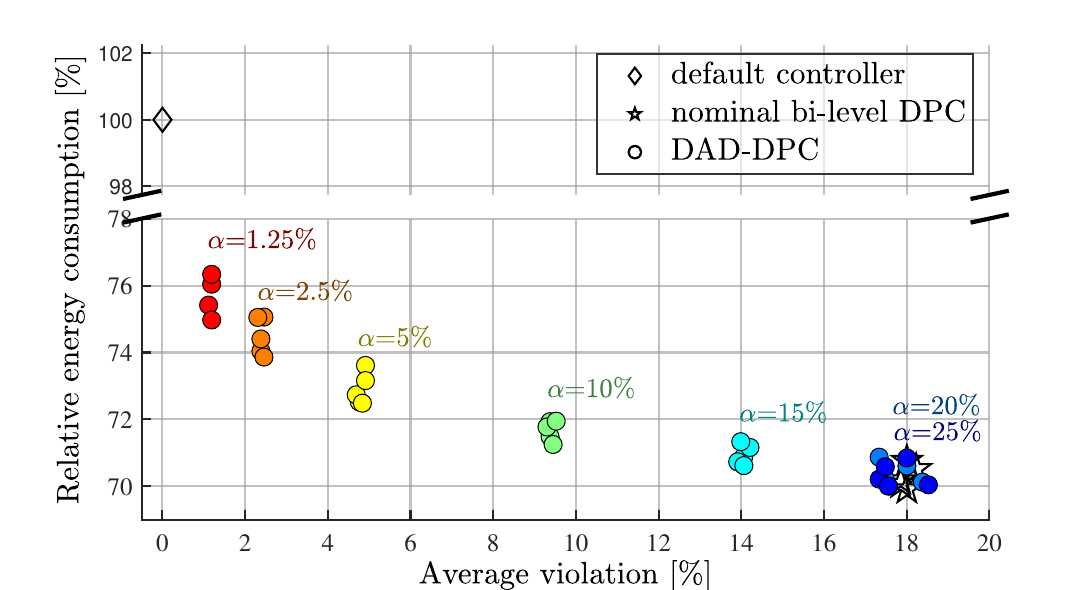}
        \caption{Case 3 during peak heating days: the relative energy consumption and average comfort violations of different controllers. For DAD-DPC, each color corresponds to a different $\alpha$}
    \label{fig:case3_scatter}
\end{figure}

To evaluate the trade-off between violations and energy consumption, we tested the DAD-DPC with more $\alpha$ values and compared its performance with the default controller and the nominal DPC (\change{i.e., $\pi(z_t, \mathcal{D}(1))$, corresponding to $d_{i} \in \mathcal{D}_i(\sigma)=\{\mathbf{0}\}$ in the RB-DPC~\eqref{eqn:dpc_robust}}).  For all controllers except the default, five Monte Carlo simulations were conducted using different realizations of measurement noise. Figures~\ref{fig:case1_scatter}, \ref{fig:case2_scatter} and~\ref{fig:case3_scatter} display scatter plots of total relative energy consumption versus average violation level. Each marker shape represents a specific controller, and each data point corresponds to one Monte Carlo trial. For the DAD-DPC controller, each color indicates a different choice of $\alpha$.
% In scenarios with $\alpha=0\%$ across all three cases, the RB-DPC  $\pi(z_t,\mathcal{D}(0))$ resulted in minor violation levels, indicating that Property 3 held for relatively large $\alpha$ values. Theorem~\ref{thm:bound_suf_2} guarantees that this property, as part of the sufficient conditions, ensures satisfaction of the asymptotic bound~\eqref{eqn:cons_vio_asy}. These findings were consistent with the results of the other scenarios ($\alpha \geq 1.25\%$).
% across different measurement noise scenarios
Table~\ref{tab:energy_violation_case1} summarizes the Monte Carlo averages of total relative energy consumption, average violation levels, \change{reported in both ratio and magnitude,} for each $\alpha$. \change{Violation magnitude is computed as the product of the temperature deviation from the comfort bound and the sampling time, expressed in Kelvin-hours [Kh]. This metric serves as an additional comfort violation criterion, as suggested by the European indoor environment standard~\cite{comite2007indoor}.}

\begin{table*}[!ht]
\centering
\begin{threeparttable}
\caption{Monte Carlo averages of total relative energy consumption, average violations \\ \change{and computation time of DPC} for different $\alpha$ values in Cases 1, 2, 3.}
\label{tab:energy_violation_case1}
\renewcommand{\arraystretch}{1.1}
\begin{tabularx}{0.8\textwidth}{@{}M N{30mm} M M M M M M M M@{}}
\toprule
\textbf{Case} & \textbf{Type} & 
 $\alpha$ =  1.25\%& $\alpha$ =  2.5\%& $\alpha$ =  5\%& $\alpha$ =  10\%& $\alpha$ =  15\%& $\alpha$ =  20\%& $\alpha$ =  25\%&  Nominal DPC \\ 
 \midrule
\multirow{4}{*}{\textbf{1}} 
& Energy [\%]  & 71.01 & 70.51 & 69.87 & 69.15 & 68.62 & 68.22 & 68.13 & 67.95\\ \cline{2-10}
& Violation \change{ratio} [\%] & 1.10 & 2.23 & 4.46 & 9.12 & 13.76 & 17.43 & 18.51 & 19.93 \\ 
\cline{2-10} & \change{Violation magnitude [Kh]} & \change{0.00} & \change{0.01} & \change{0.03} & \change{0.34} & \change{1.07} & \change{1.90} & \change{2.25} & \change{2.42} \\
\cline{2-10} &  \change{Computation time [ms]} & \change{6.03} & \change{6.10} & \change{6.19} & \change{6.19} & \change{6.33} & \change{6.25} & \change{6.32} & \change{6.18} \\
\midrule
\multirow{4}{*}{\textbf{2}} 
& Energy [\%]  & 91.10 & 89.92 & 88.77 & 87.29 & 86.66 & 86.07 & 85.80 & 85.58 \\ \cline{2-10}
& Violation \change{ratio} [\%]  & 1.07 & 2.13 & 4.24 & 8.26 & 11.64 & 14.96 & 18.08 & 20.79 \\
\cline{2-10} & \change{Violation magnitude [Kh]} & \change{0.25} & \change{0.44} & \change{0.88} & \change{1.77} & \change{2.53} & \change{3.68} & \change{5.40}   & \change{7.05} \\
\cline{2-10} &  \change{Computation time [ms]} & \change{6.17} & \change{6.15} & \change{6.22} & \change{6.19} & \change{6.25} & \change{6.27} & \change{6.30} & \change{6.33} \\
\midrule
\multirow{4}{*}{\textbf{3}} 
& Energy [\%]  & 77.03 & 74.48 & 72.90 & 71.68 & 70.95 & 70.36 & 70.34 & 70.29\\ 
\cline{2-10} & Violation \change{ratio} [\%] & 1.19 & 2.40 & 4.82 & 9.40 & 14.05 & 17.75 & 17.78 & 17.99 \\ 
\cline{2-10} & \change{Violation magnitude [Kh]} & \change{0.15} & \change{0.29} & \change{0.65} & \change{1.59} & \change{2.42} & \change{3.44} & \change{3.52} & \change{3.49} \\
\cline{2-10} &  \change{Computation time [ms]} & \change{11.39} & \change{11.69} & \change{11.56} & \change{11.43} & \change{11.23} & \change{11.23} & \change{11.22} & \change{10.73} \\
\bottomrule
\end{tabularx}
\begin{tablenotes}
\scriptsize
\item \hfill \change{Violation magnitude = temperature deviation $\times$ sampling time (in Kelvin-hours [Kh]}).
\end{tablenotes}
\end{threeparttable}
\end{table*}
% case 2
% & 89.56 & 88.40 & 87.50 & 86.70 & 86.37 & 86.07 & 85.80 & 85.58
% & 1.15 & 2.14 & 4.26 & 8.32 & 11.65 & 14.94 & 18.05 & 20.79 
% dpc: 85.5835,20.79%, 5%: 88.7680,4.24%
% case 3
% & 75.15  & 73.98 & 72.93 & 71.68 & 71.02 & 70.30 & 70.29 & 70.29
% & 1.25 & 2.49 & 4.81 & 9.43 & 14.08 & 17.74 & 17.99 & 17.99
% dpc: 70.2915,17.99%, 5%: 72.9031,4.82%

The above results clearly illustrate the trade-off between comfort violations  \change{(both in ratio and magnitude)} and energy consumption: lower $\alpha$ values reduced average violations and increased energy consumption. 
Compared to the conservative default controller for Cases 1, 2 and 3, DAD-DPC with a 5\% average violation level, aligned with the European standard EN15251 Annex G~\cite{comite2007indoor}, reduced energy consumption by \textbf{30.1\%/11.2\%/27.1\%}, respectively, demonstrating significant efficiency improvements.
Furthermore, when compared to the nominal DPC, which let to more violations, the 5\%-violation DAD-DPC setting resulted in only a marginal increase in energy consumption (\textbf{2.8\%/3.7\%/3.7\%})   while achieving significantly fewer violations (\textbf{77.6\%/79.6\%/73.2\%}) across the three cases. These findings highlight the ability of DAD-DPC to balance comfort and energy efficiency effectively.

\change{While both violation ratio and magnitude showed trade-off behavior with energy consumption, they showed different patterns across the three cases.
In all cases, the violation ratio approached the specified bound $\alpha$, but the violation showed clear variation. For example, when $\alpha=5\%$, the violation ratios were consistently close to the target, but the violation magnitudes were $0.03$Kh, $0.88$Kh, and $0.65$Kh for the three cases, respectively. This is because, in the current DAD-DPC framework, the adaptive update of $\alpha_t$ in~\eqref{eqn:dad_alpha} is based on the binary violation indicator $v_t$, leading to feedback on violation ratio, as discussed in Section~\ref{sect:III_heuristic}. 
The authors of~\cite{korda2014stochastic} achieved different types of violation bounds by redefining $v_t$, assuming the disturbance distribution is known. Extending the DAD-DPC framework to account for violation magnitude using a similar approach is a promising direction for future research.}

\change{Moreover, Table~\ref{tab:energy_violation_case1} reports the average computation times for the RB-DPC $\pi(z_t,\mathcal{D}(\sigma))$ defined in~\eqref{eqn:dpc_robust} under various $\alpha$ values, and for the nominal DPC $\pi(z_t,\mathcal{D}(1))$. The results are consistent with the analysis in Remark~\ref{remark:computation}.
Across all the simulations, the optimizations were efficiently solved within milliseconds, as they were QP problems. In each building case, the RB-DPC and the nominal DPC required similar computation times, due to their equivalent computational complexity.}

Overall, this simulation study validates DAD-DPC's ability to satisfy the violation bound~\eqref{eqn:cons_vio_asy} for specified $\alpha$ levels using DAD-DPC. It underscores the potential of DAD-DPC to balance average comfort violations and energy cost savings in building climate control applications.

% goes to nomimal DPC. Q slack for more savings

% \subsubsection{Sensitivity analysis of \texorpdfstring{$\eta$}{eta}}

% Sensitivity analysis

% Analysis: different eta, different behavior, but still convergence

\section{Case studies: experiments} \label{sect:experiment}
This section presents experimental studies conducted to validate the efficacy of the proposed DAD-DPC framework. 

\subsection{Setup in an occupied campus building}

% \vspace{1em}

\begin{figure}[!ht]
    \centering
    \includegraphics[width=0.9\linewidth]{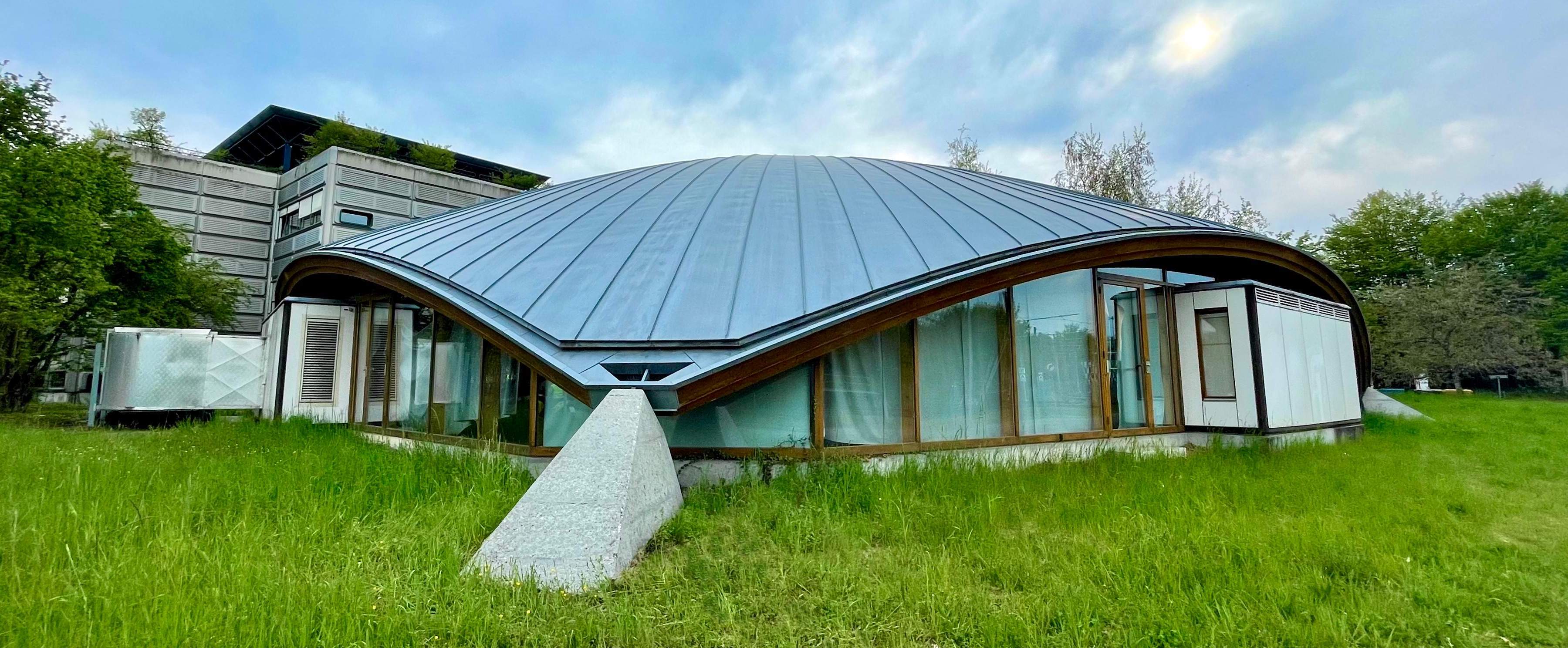}
        \caption{Photo of the Polydome.}
    \label{fig:polydome}
\end{figure}

\begin{figure}[!ht]
    \centering
    \includegraphics[width=1\linewidth]{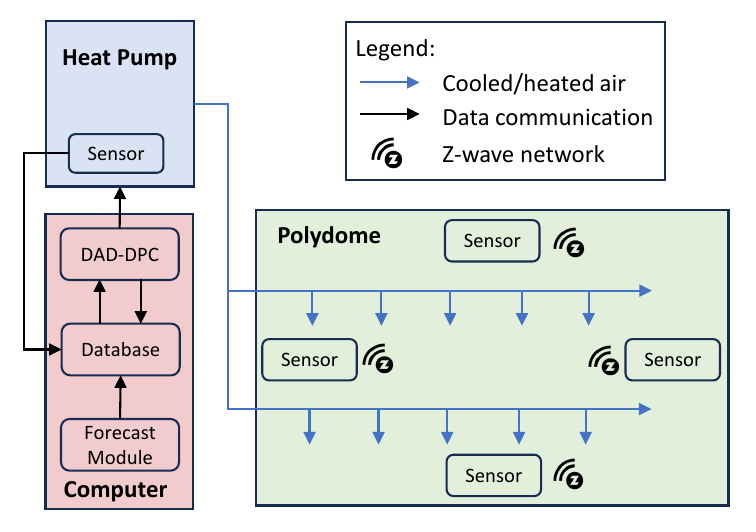}
        \caption{The overall system layout of the Polydome. Information is exchanged through wired communications and a wireless Z-wave network.}
    \label{fig:polydome_concept}
\end{figure}

The experiments were conducted in an occupied campus building, called Polydome (Figure~\ref{fig:polydome}). The Polydome is a standalone building of 600 $m^2$,  functioning as a lecture hall, exam venue and conference room with a seating capacity of 200 people. The system layout for this case study is depicted in Figure~\ref{fig:polydome_concept}, where blue arrows represent the flow of cooled or heated air, and black arrows indicate the direction of data communication. The main components are detailed below.

% HP, Polydome (Sensors), Computer (Database, Prediction Module, DPC)

\textbf{\ac{hp}}: An AERMEC RTY-04 HP is installed in the Polydome. The HP operates in either heating or cooling mode.  Its ventilation fan distributes ambient, cooled, or heated air throughout the building via two elongated fabric ducts, as shown by the blue arrows in Figure~\ref{fig:polydome_concept}.
The nominal capacities of the ventilation fan, cooling-mode compressor, and heating-mode compressor are $2.4 \, kW$, $4.6 \, kW$ and $6 \, kW$ respectively.

\textbf{Computer}: A central computer (iMac with 32 GB 1600 MHz DDR3 memory and a 3.4 GHz Intel Core i7) manages several components. (1) Database:  Data, including HP power consumption, indoor temperature, and historical weather data, are stored in a time-series database, InfluxDB 1.3.7~\cite{influxdb}. (2) DAD-DPC: The DAD-DPC framework was implemented in MATLAB. It calculated the control variables and communicated them to the HP using the Modbus serial communication protocol~\cite{swales1999open}. (3) Prediction Module: Weather forecasts, including outdoor air temperature and solar radiation, were obtained via the Tomorrow.io~\cite{tomorrow} API. Forecast data and current weather conditions were updated every 15 minutes.

\textbf{Sensors}: The HP's power consumption was monitored using an EMU 3-phase meter~\cite{emu}, which transmitted data to the central computer via Ethernet.  Indoor temperature was monitored using four Fibaro Door/Window Sensor v2 units, installed at a height of 1.5 meters on different walls (locations shown in Figure~\ref{fig:polydome_concept}). These sensors transmitted temperature data to the central computer every five minutes using the Z-wave wireless protocol. For this study, the building’s output, $y$, was calculated as the average temperature measured by the four sensors.

\subsection{Setup of controllers} 

\subsubsection{Default controller}

The default controller implemented in the Polydome’s BMS is described here. 
(1) A scheduler operates the fan continuously to ensure ventilation and determines whether the HP operates in heating or cooling mode based on outdoor conditions. Access to this scheduler was \textbf{not} authorized for this study.
(2) A bang-bang controller regulates the indoor temperature based on a setpoint and the return-air temperature, using a one $^\circ C$ deadband. For instance, in cooling mode, if the return-air temperature falls one $^\circ C$ below the setpoint, the HP activates at full power until the return-air temperature reaches the setpoint. The setpoints are fixed at ${21}^\circ C$ for the heating mode and ${23}^\circ C$ for the cooling mode.

% This default controller was used as a low-level controller to execute the optimal input computed by DAD-DPC during the experiments. 
\change{In the DAD-DPC implementation, this default} bang-bang control policy was used as \change{$\pi^B$, with $\mathcal{Y}_{\text{lim}}$ chosen as a suitable operating range, $\{y|{15}^\circ C \leq y \leq {30}^\circ C \}$.}
% Additionally, data from historical days when the default controller directly operated the system are used as a baseline for comparative analysis in Section~\ref{sect:exp_result}. 

\subsubsection{Setup of DAD-DPC}

% Building (y), HVAC (u). Describe the input and output measurement, time step.  Low-level controllers

In the experiment, DAD-DPC was implemented for temperature control, while the HP’s ventilation and operating mode (heating or cooling) followed the default scheduler.

The \ac{io} configuration for DAD-DPC was identical to Case 1 in the simulation study due to the similar HVAC setup, shown in Table~\ref{tab:simu_dpc_io}\change{, i.e., using HP electrical power as input and indoor temperature as output}. Since the HP operated exclusively in either heating or cooling mode, separate \ac{io} datasets were collected for each mode to develop the corresponding RB-DPC and SCP estimators. 
After the DAD-DPC computed the optimal HP electrical power, modulation similar to Case 1 was used to control the HP \change{using the default controller as the low-level controller}. \change{Within each sampling
period}, by adjusting the default controller\change{'s setpoint between high and low values}, the HP alternated between on and off states. The on/off durations were calculated \change{such that the resulting average electrical power matched the optimal value computed by
DAD-DPC. The actual average electrical power within each sampling period was used as the input measurement.}  

The parameters for the DAD-DPC method were chosen as follows: sampling time, 15 minutes; prediction horizon $N$, 24 steps; initial steps $t_{init}$, 12 steps; regularization weight $Q_g$, 0.01$I$; regularization weight $Q_{\delta}$, 10.  
\change{The sampling time and prediction horizon were selected based on our previous control experience in the Polydome. They are also consistent with the practical MPC guidelines summarized in Section IV.B.2). The choice of $N=24$ (6 hours) was further constrained by the weather forecast API, Tomorrow.io~\cite{tomorrow}, provided up to a 6-hour prediction in the 15-min sampling time.}
The segmented trajectory trick from~\cite{o2022data} was employed to reduce the data requirement for Hankel matrices. In the RB-DPC, the objective was defined as  $I(y_t,u_t)= u_t$ and the input constraint $\mathcal{U}$ was set to the latest maximal electrical power. 
\change{Under these settings, the resulting RB-DPC was a QP problem as analyzed in Remark 2, and was solved by MATLAB Quadprog~\cite{matlab:quadprog}.}

The experimental study spanned over two months and included various operational settings. The comfort constraints and parameters ($\alpha, \alpha_{0}$, $\eta$) were adjusted as described in Section~\ref{sect:exp_result}. To address the nonlinearity in the HP’s coefficient of performance, adaptive updates were incorporated into the RB-DPC~\cite{shi2025adaptive}, and residual updates were applied to SCP for long-term operations. Initially, both the Hankel matrix and SCP data lengths were set to 480 steps (5 days). The required 10-day datasets for each mode were collected using the default controller.

\subsection{Results of experiments} \label{sect:exp_result}

\begin{figure*}[!ht]
    \centering
    \includegraphics[width=1.0\linewidth]{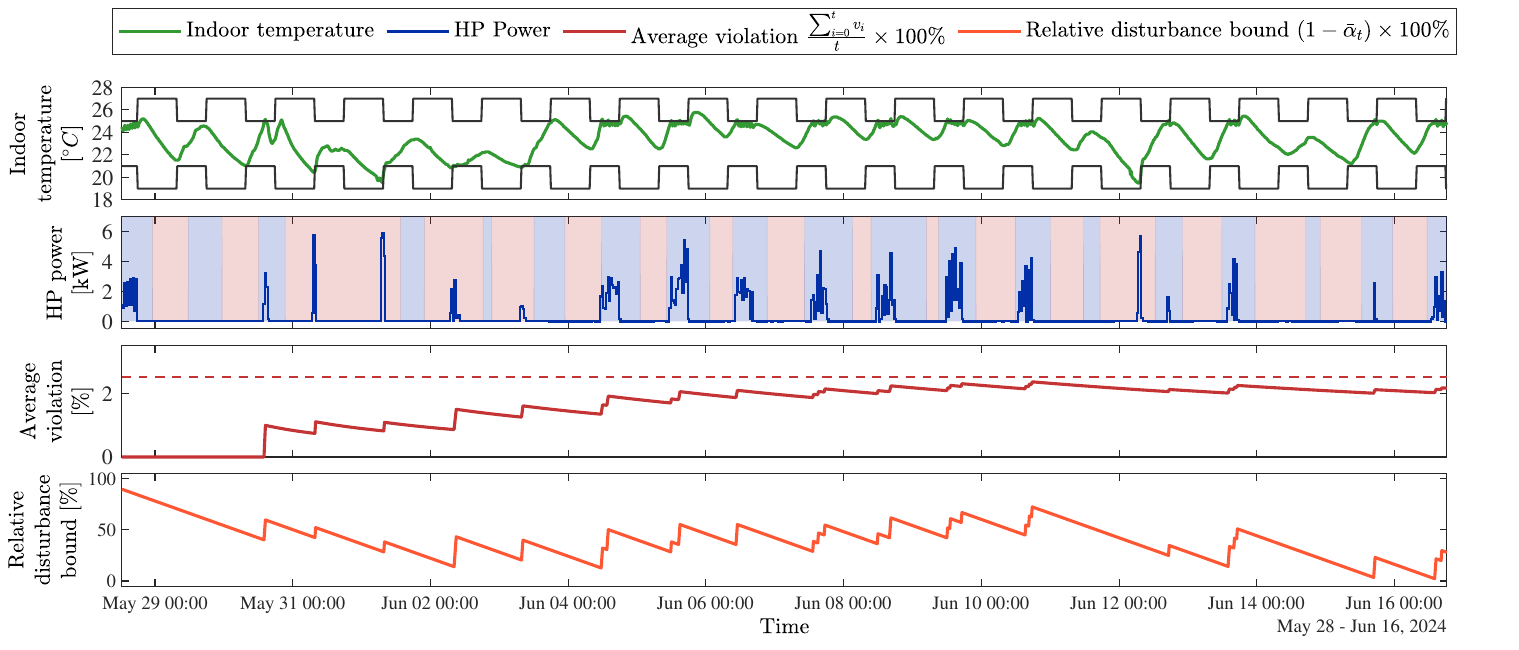}
        \caption{Scenario 1 from May 28$^{\text{th}}$ to June 16$^{\text{th}}$, $\alpha=2.5\%$. Sub-figures from top to bottom include: indoor temperature; HP's electrical power (blue means cooling mode and red means heating mode in the background); average violations (the dashed line indicates the $\alpha$ bound); relative disturbance bound.}
    \label{fig:exp_MayJune}
    % \vspace{1em}
\end{figure*}

\begin{figure}[!ht]
    \centering
    \includegraphics[width=1.0\linewidth]{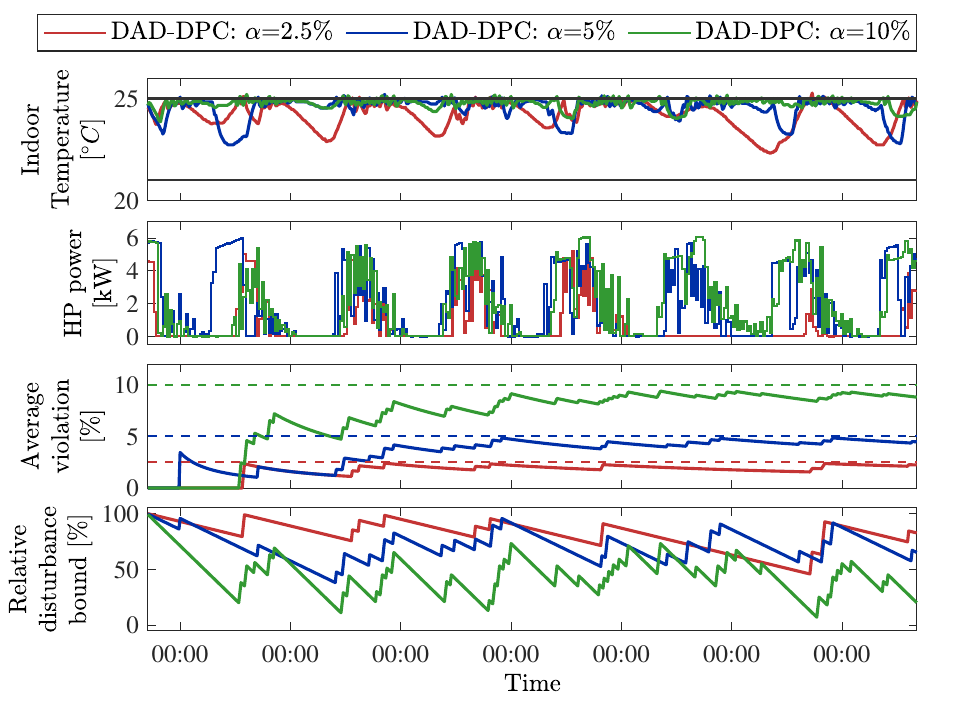}
        \caption{Three scenarios:  Scenario 2 from August 20$^{\text{th}}$ to 27$^{\text{th}}$, $\alpha=2.5\%$; Scenario 3, from July 30$^{\text{th}}$ to August 6$^{\text{th}}$, $\alpha=5\%$; Scenario 4 from August 6$^{\text{th}}$ to 13$^{\text{rd}}$, $\alpha=10\%$;
        Sub-figures from top to bottom include: indoor temperature; HP's electrical power (always in cooling mode); average violations (the dashed lines indicate three $\alpha$ bounds); relative disturbance bound.}
    \label{fig:exp_3daddpc}
    % \vspace{1em}
\end{figure}

This section presents the results from four DAD-DPC operational scenarios spanning a total of 40 days, designed to validate the framework’s efficacy. Additional scenarios, covering 36 days, are included in Appendix~\ref{appen:exp_result} as they provide less critical insights. 
\change{Table~\ref{tab:scenarios_main} summarizes the settings for the four main scenarios.}

\begin{table}[!ht]
{\color{black}
\centering
\caption{Main experimental scenarios}
\label{tab:scenarios_main}
\renewcommand{\arraystretch}{0.2}
\setlength{\tabcolsep}{4pt}
\begin{tabularx}{\columnwidth}{@{}C{10mm} p{19mm} C{5mm} Y C{4mm} C{3mm} c@{}}
\toprule
\textbf{Scenario} & \textbf{Dates (2024)} & \textbf{Days} & \textbf{Comfort bounds} & $\boldsymbol{\alpha}$ & $\boldsymbol{\alpha_0}$ & $\boldsymbol{\eta}$ \\
\midrule
1 & May 28--Jun 16 & 19 & 08{:}00--18{:}00: 21--25\,\textdegree C; otherwise: 19--27\,\textdegree C & 2.5\% & 0.1 & 0.1 \\ \midrule
2 & Aug 20--Aug 27 & 7  & All times: 21--25\,\textdegree C & 2.5\% & 0   & 0.1 \\ \midrule
3 & Jul 30--Aug 06 & 7  & All times: 21--25\,\textdegree C & 5\%   & 0   & 0.1 \\ \midrule
4 & Aug 06--Aug 13 & 7  & All times: 21--25\,\textdegree C & 10\%  & 0   & 0.1 \\ 
\bottomrule
\end{tabularx}
}
\end{table}

% \begin{itemize} [nolistsep,leftmargin=1.5em]
%     \item \textbf{Scenario 1.} Duration: 19 days from May 28$^{\text{th}}$ to June 16$^{\text{th}}$, 2024;  Parameters: $21^\circ C \leq y_t 
%  \leq 25^\circ C$ from 8 a.m. to 6 p.m., $19^\circ C \leq y_t 
%  \leq 27^\circ C$ otherwise, $\alpha=2.5\%$, $\alpha_0=0.1$, $\eta=0.1$.
%     \item \textbf{Scenario 2.} Duration: 7 days from August 20$^{\text{th}}$ to 27$^{\text{th}}$, 2024; Parameters: $21^\circ C \leq y_t 
%  \leq 25^\circ C$ at any time, $\alpha=2.5\%$, $\alpha_0=0$, $\eta=0.1$.
%     \item \textbf{Scenario 3.} Duration: 7 days from July 30$^{\text{th}}$ to August 6$^{\text{th}}$, 2024; Parameters: $\alpha=5\%$, others are the same as Scenario 2.
%     \item \textbf{Scenario 4.} Duration: 7 days from August 6$^{\text{th}}$ to 13$^{\text{rd}}$, 2024; Parameters: $\alpha=10\%$, others are the same as Scenario 2.    
% \end{itemize}

Figure~\ref{fig:exp_MayJune} presents the results for Scenario 1, which corresponds to the transitional period between spring and summer. During this time, the HP frequently alternated between heating and cooling modes due to varying outdoor temperatures.

The first two sub-figures in Figure~\ref{fig:exp_MayJune} depict the indoor temperature and the HP’s power consumption trajectories. The DAD-DPC effectively maintained the indoor temperature within the specified comfort constraints for most of the period, even with frequent mode switching between heating and cooling. 
The third and fourth sub-figures depict the trajectories of the average violation levels ($\frac{\sum_{i=0}^{t} v_i}{t}\times 100\%$) and the adaptive disturbance bound updates (($(1-\bar{\alpha}_t)\times 100\%$)). The average violation level \change{satisfied the specified bound $\alpha=2.5\%$ asymptotically}, demonstrating the effectiveness of the DAD-DPC framework. The well-designed DAD-DPC dynamically adjusted the disturbance bound in response to violations, ensuring that $\alpha_t$ remained bounded. \change{As a result, indicated by Lemma~\ref{thm:bound_suf_1}, the average violation converged to the specified $\alpha=2.5\%$ after 19 days of operation, aligning with thermodynamic intuition.} With a small $\alpha_0=0$, the strict bound~\eqref{eqn:cons_vio_rob} was also satisfied.

% It empirically validated that a well-designed DAD-DPC with Property 1 can result in the sufficient condition indicated by Lemma~\ref{thm:bound_suf_1}. 

% One interesting observation is that, as weather varied within this scenario, the relative disturbance bound changed in a large range, i.e. around  0\% to 80\%. It highlights the necessity of adaptive disturbance updates to achieve a specified violation level in building climate control.

Scenarios 2, 3 and 4 were conducted during the summer to evaluate the performance of DAD-DPC under different violation bounds, i.e. $\alpha \in \{ 2.5\%, 5\%, 10\% \}$. All the other parameters were the same, and the constant comfort constraint was applied to ensure a fair comparison with the default controller. The closed-loop trajectories for these scenarios are shown in Figure~\ref{fig:exp_3daddpc}, with distinct colors representing different settings. Since these scenarios occurred during the summer in Lausanne, the HP primarily operated in cooling mode, and the indoor temperature often approached the upper comfort bound. Direct comparison of input and output behaviors across different $\alpha$ values is challenging due to varying weather conditions. For instance, in Scenario 2, cooler weather occasionally caused the indoor temperature to drop to 22$^\circ$. In the fourth sub-figure, the relative disturbance bound is shown to adaptively update, with smaller target $\alpha$ values resulting in larger bounds to ensure fewer violations. Consequently, the average violations in each scenario converged to their respective chosen bounds, satisfying the violation constraint~\eqref{eqn:cons_vio_asy}, as seen in the third sub-figure.

% It can be observed in the top two sub-figures in Figure~\ref{fig:exp_3daddpc}: the indoor temperature decreased without any cooling power at night.

\begin{table*}[!ht]
\centering
\begin{threeparttable}
\caption{Comparison of scenarios with different controllers under similar weather conditions.} 
\label{tab:controller_comparison}
\renewcommand{\arraystretch}{0.2} \setlength{\tabcolsep}{4pt} 
\begin{tabularx}{0.78\textwidth}{@{}c c c c c c@{}}
\toprule
 & \makecell{\change{Average Violation} \\ \change{Ratio [\%]} } & \makecell{\change{Average Violation} \\ \change{Magnitude [Kh]} } & \makecell{Average Hourly Energy \\  Consumption [kWh]} & \makecell{Average  Outdoor \\ Temperature  [$^\circ$C]} & \makecell{Average Solar \\ Radiation  [kW$/$m$^2$]} \\ \midrule
Scenario 3     & 5.00   & \change{0.48}      & 1.67        & 23.11       & 0.26  \\ \midrule
Scenario 4     & 8.33   & \change{0.72}       & 1.59       & 23.09      & 0.25  \\ \midrule
Default    & 0.00       & \change{0.00}  & 2.10        & 23.09       & 0.22  \\ 
\bottomrule
\end{tabularx}
\begin{tablenotes}
\scriptsize
\item \hfill \change{Violation magnitude = temperature deviation $\times$ sampling time (in Kelvin-hours [Kh]}).
\end{tablenotes}
\end{threeparttable}
\end{table*}

% \begin{table*}[!ht]
% \centering
% \begin{threeparttable}
% \caption{Comparison of scenarios with different controllers under similar weather conditions.} 
% % \label{tab:controller_comparison}
% \begin{tabular}{|c|c|c|c|c|c|}
% \hline
%  & \makecell{Average Violation \\ Ratio [\%] } & \makecell{\change{Average Violation} \\ \change{Magnitude [Kh]} } & \makecell{Average Hourly Energy \\  Consumption [kWh]} & \makecell{Average  Outdoor \\ Temperature  [$^\circ$C]} & \makecell{Average Solar \\ Radiation  [kW$/$m$^2$]} \\ \hline
% Scenario 3     & 5.00   & \change{0.48}      & 1.67        & 23.11       & 0.26  \\ \hline
% Scenario 4     & 8.33   & \change{0.72}       & 1.59       & 23.09      & 0.25  \\ \hline
% Default    & 0.00       & \change{0}  & 2.10        & 23.09       & 0.22  \\ \hline
% \end{tabular}
% \begin{tablenotes}
% \scriptsize
% \item \hfill \change{Violation magnitude = temperature deviation $\times$ sampling time (in Kelvin-hours [Kh]}).
% \end{tablenotes}
% \end{threeparttable}
% \end{table*}

To compare the violations and energy consumption under different $\alpha$ settings, continuous days with similar weather conditions were selected: Scenario 3 (from July 31$^{\text{st}}$ to August 5$^{\text{th}}$), Scenario 4 (from August 6$^{\text{th}}$ to 10$^{\text{th}}$), and the default controller (from August 29$^{\text{th}}$ to September 1$^{\text{st}}$). 
The statistical results are summarized in Table~\ref{tab:controller_comparison}, \change{which} empirically demonstrated the trade-off between comfort violations \change{(both in ratio and magnitude)} and energy savings enabled by DAD-DPC.
Scenario 2 was excluded from this comparison due to cooler weather.  Scenarios 3 and 4 demonstrated substantial energy savings, reducing consumption by \textbf{20.5\%} and \textbf{24.3\%}, respectively, compared to the default controller. Furthermore, Scenario 3 increased energy consumption by only \textbf{5.0\%} compared to Scenario 4 while achieving \textbf{39.8\%} fewer comfort violations.

\section{Conclusion}
This paper proposes a DAD-DPC framework designed to ensure the asymptotic average violation bound. As a practical application in building climate control, we developed a data-driven implementation of DAD-DPC utilizing Willems' Fundamental Lemma and conformal prediction.
The framework's performance was evaluated through high-fidelity simulations on the BOPTEST platform and experiments in the occupied campus building, Polydome. 
Compared to the conservative default controller in four cases, DAD-DPC with a 5\% average violation level reduced energy consumption by 30.1\%/11.2\%/27.1\%/20.5\%, demonstrating significant efficiency improvements.
Additionally, the 5\%-violation DAD-DPC framework reduced comfort violations by 77.6\%/79.6\%/73.2\%/39.8\% while increasing energy consumption by only 2.8\%/3.7\%/3.7\%/5.0\% compared to more relaxed violation-bound configurations. These results underscore the framework's capability to balance energy efficiency and comfort effectively, making it a promising solution for real-world building climate control applications.

\appendix

\subsection{Other experiment results} \label{appen:exp_result}

This appendix describes additional experimental scenarios conducted in the Polydome, spanning a total of 36 days.
\change{Table~\ref{tab:scenarios_app} summarizes their settings.}

\begin{table}[!ht]
{\color{black}
\centering
\caption{Additional experimental scenarios}
\label{tab:scenarios_app}
\renewcommand{\arraystretch}{0.2}
\setlength{\tabcolsep}{4pt}
\begin{tabularx}{\columnwidth}{@{}C{9mm} p{19mm} C{4.5mm} Y C{4mm} C{3mm} c@{}}
\toprule
\textbf{Scenario} & \textbf{Dates (2024)} & \textbf{Days} & \textbf{Comfort bounds} & $\boldsymbol{\alpha}$ & $\boldsymbol{\alpha_0}$ & $\boldsymbol{\eta}$ \\
\midrule
5 & Aug 13--Aug 20 & 7  & All times: 21--25\,\textdegree C & 5\%   & 1.0 & 0.1 \\ \midrule
6 & May 14--May 28 & 14 & 08{:}00--18{:}00: 21--25\,\textdegree C; otherwise: 19--27\,\textdegree C & 5\% & 0.1 & 0.03 \\ \midrule
7 & Jul 08--Jul 15 & 8  & 08{:}00--18{:}00: 21--25\,\textdegree C; otherwise: 19--27\,\textdegree C & 2.5\% & 0.1 & 0.1 \\ \midrule
8 & Jul 17--Jul 24 & 7  & 08{:}00--18{:}00: 21--25\,\textdegree C; otherwise: 19--27\,\textdegree C & 2.5\% & 0   & 0.1 \\ 
\bottomrule
\end{tabularx}
}
\end{table}

% \begin{itemize} [nolistsep,leftmargin=1.5em]
%     \item \textbf{Scenario 5.} Duration: 7 days from August 13$^{\text{th}}$ to 20$^{\text{th}}$, 2024; Parameters: $21^\circ C \leq y_t 
%  \leq 25^\circ C$ at any time, $\alpha=5\%$, $\alpha_0=1$, $\eta=0.1$. 
%     \item \textbf{Scenario 6.} Duration: 14 days from May 14$^{\text{th}}$ to 28$^{\text{th}}$, 2024;  Parameters: $21^\circ C \leq y_t 
%  \leq 25^\circ C$ from 8 a.m. to 6 p.m., $19^\circ C \leq y_t 
%  \leq 27^\circ C$ otherwise, $\alpha=5\%$, $\alpha_0=0.1$, $\eta=0.03$.  
%  \item \textbf{Scenario 7.} Duration: 8 days from July 8$^{\text{th}}$ to 15$^{\text{th}}$, 2024; Parameters: $21^\circ C \leq y_t 
%  \leq 25^\circ C$ from 8 a.m. to 6 p.m., $19^\circ C \leq y_t 
%  \leq 27^\circ C$ otherwise, $\alpha=2.5\%$, $\alpha_0=0.1$, $\eta=0.1$.
%  \item \textbf{Scenario 8.} Duration: 7 days from July 17$^{\text{th}}$ to 24$^{\text{th}}$, 2024; Parameters: $21^\circ C \leq y_t 
%  \leq 25^\circ C$ from 8 a.m. to 6 p.m., $19^\circ C \leq y_t 
%  \leq 27^\circ C$ otherwise, $\alpha=2.5\%$, $\alpha_0=0$, $\eta=0.1$.
% \end{itemize}

\begin{figure}[!ht]
    \centering
    \includegraphics[width=1.0\linewidth]{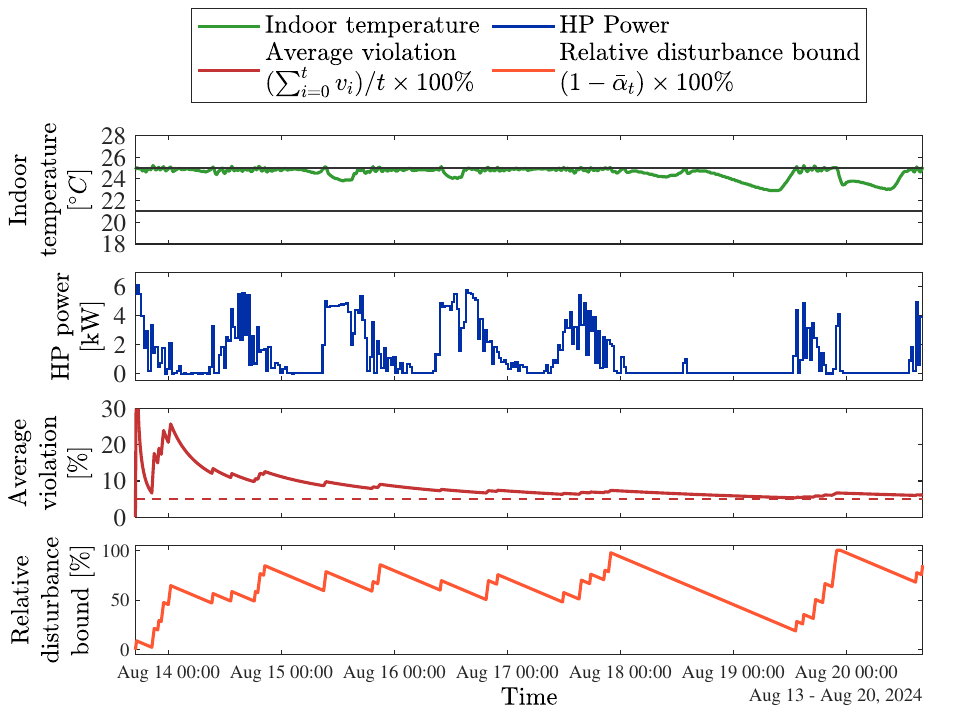}
        \caption{ Scenario 5 from August 13$^{\text{th}}$ to 20$^{\text{th}}$, $\alpha=5\%$.
        Sub-figures from top to bottom include: indoor temperature; HP's electrical power (always in cooling mode); average violations (the dashed line indicates the $\alpha$ bound); relative disturbance bound.}
    \label{fig:exp_apen_F1Aug}
    % \vspace{1em}
\end{figure}

% Appendix  (1) different $\alpha_0$ 0.05 
Scenario 5 used the same parameter as Scenario 3 in Section~\ref{sect:exp_result}, except for a larger initial $\alpha_0=1$. As shown in Figure~\ref{fig:exp_apen_F1Aug}, the average violation ultimately converged to satisfy the asymptotic bound~\eqref{eqn:cons_vio_asy}. During the operation, the average violation was higher than $\alpha$, consistent with the behavior indicated by~\eqref{eqn:thm_bound} due to the large $\alpha_0$.

\begin{figure}[!ht]
    \centering
    \includegraphics[width=1.0\linewidth]{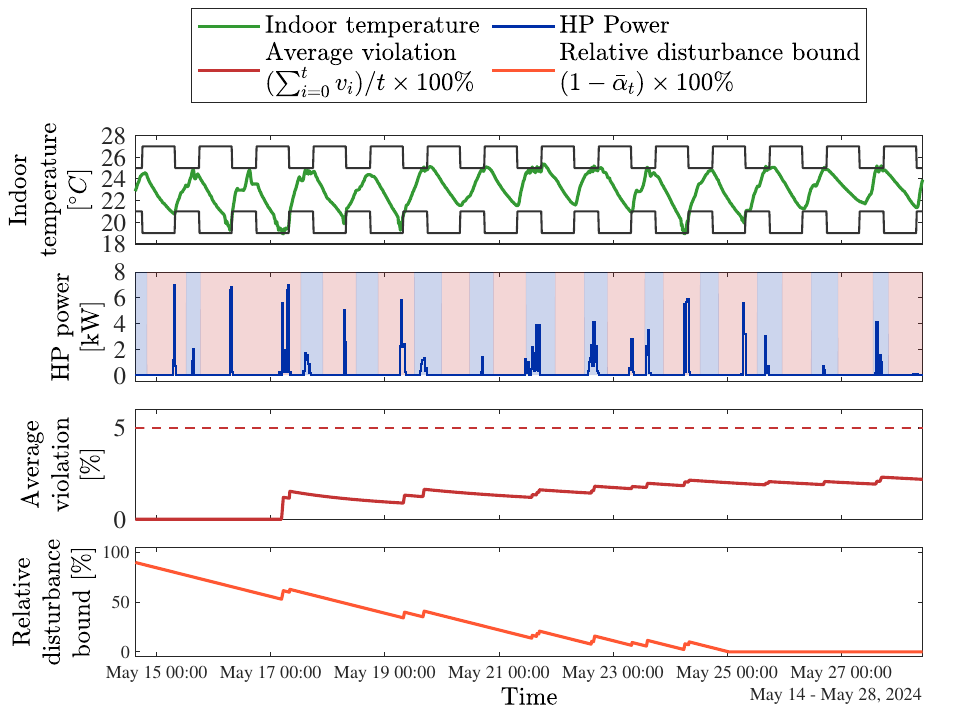}
        \caption{Scenario 6 from May 14$^{\text{th}}$ to 28$^{\text{th}}$, $\alpha=5\%$.
        Sub-figures from top to bottom include: indoor temperature; HP's electrical power (blue means cooling mode and red means heating mode in the background); average violations (the dashed line indicates the $\alpha$ bound); relative disturbance bound.}
    \label{fig:exp_apen_F2May}
    % \vspace{1em}
\end{figure}

% (2) 0.05 when mild weather (nominal DPC behavior) 
As shown in Figure~\ref{fig:exp_apen_F2May}, Scenario 6 occurred during a transition period. 
\change{Due to mild weather conditions, comfort was maintained with low energy consumption resulting in violations below $\alpha = 5\%$.} The system eventually operated as a nominal DPC $\pi(z_t, \mathcal{D}(1))$. \change{This further illustrates that enforcing violations to converge to $\alpha = 5\%$ is unnecessary in building climate control, as doing so would increase energy consumption in such scenarios.}
% As shown in Figure~\ref{fig:exp_apen_F2May}, Scenario 6 occurred during a mild transition period, similar to Scenario 1 in Section~\ref{sect:exp_result}. Due to the relatively high violation bound $\alpha=5\%$, the system eventually operated as a nominal DPC $\pi(z_t, \mathcal{D}(1))$. The adaptive updates progressed slowly because of the small update factor $\eta=0.03$, which also affected the convergence speed, as indicated by~\eqref{eqn:thm_bound}.

\begin{figure}[!ht]
    \centering
    \includegraphics[width=1.0\linewidth]{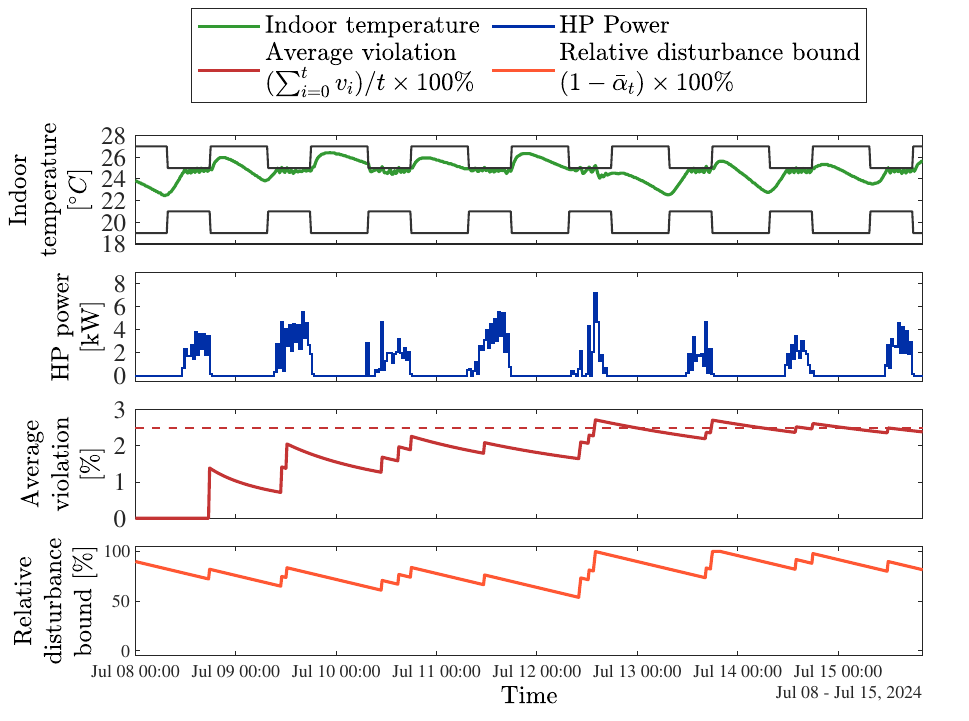}
        \caption{ 
        Scenario 7 from July 8$^{\text{th}}$ to 15$^{\text{th}}$, $\alpha=2.5\%$.
        Sub-figures from top to bottom include: indoor temperature; HP's electrical power (always in cooling mode); average violations (the dashed line indicates the $\alpha$ bound); relative disturbance bound.}
    \label{fig:exp_apen_F3JulyU2}
    % \vspace{1em}
\end{figure}

\begin{figure}[!ht]
    \centering
    \includegraphics[width=1.0\linewidth]{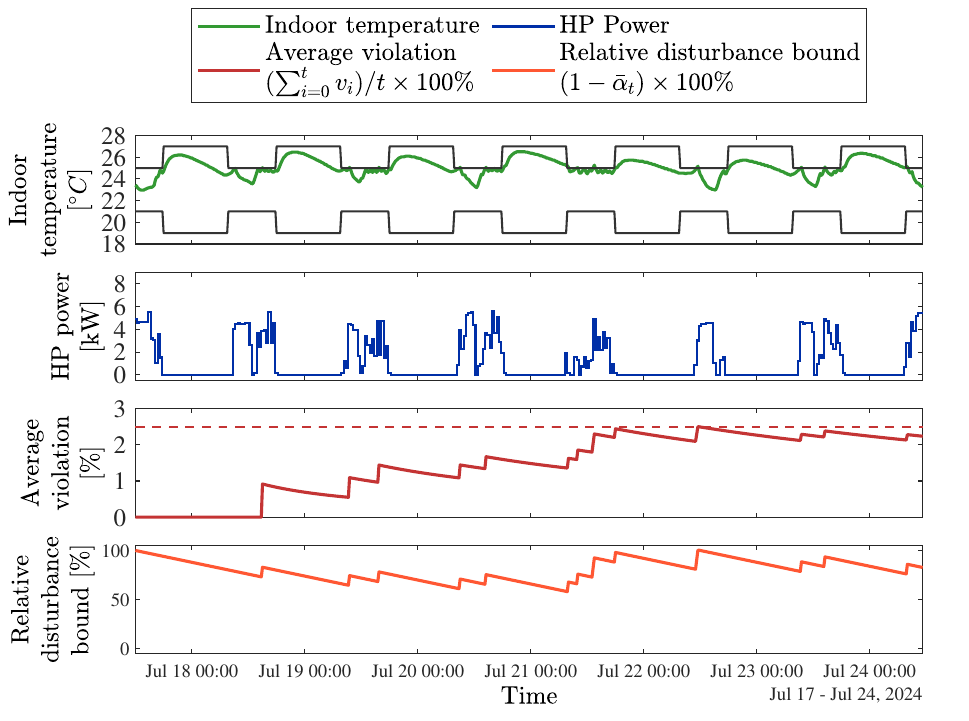}
        \caption{Scenario 8 from July 17$^{\text{th}}$ to 24$^{\text{th}}$, $\alpha=2.5\%$.
        Sub-figures from top to bottom include: indoor temperature; HP's electrical power (always in cooling mode); average violations (the dashed line indicates the $\alpha$ bound); relative disturbance bound.}
    \label{fig:exp_apen_F3JulyU1}
    % \vspace{1em}
\end{figure}

% (3) 2 0.025 when hotter, 1 or 2 compressors
Scenarios 7 and 8, which shared similar parameters, are illustrated in Figures~\ref{fig:exp_apen_F3JulyU2} and~\ref{fig:exp_apen_F3JulyU1}. However, Scenario 7 utilized two compressors in the HP, which allowed for increased input power and a larger input constraint. Consequently, the indoor temperature in Scenario 7 remained closer to the upper comfort bound, reflecting less conservative control compared to Scenario 8.   

\bibliographystyle{ieeetr}
\bibliography{ref.bib}

@article{salsbury2005survey,
  title={A survey of control technologies in the building automation industry},
  author={Salsbury, Timothy I},
  journal={IFAC Proceedings Volumes},
  volume={38},
  number={1},
  pages={90--100},
  year={2005},
  publisher={Elsevier}
}

@article{oldewurtel2012use,
  title={Use of model predictive control and weather forecasts for energy efficient building climate control},
  author={Oldewurtel, Frauke and Parisio, Alessandra and Jones, Colin N and Gyalistras, Dimitrios and Gwerder, Markus and Stauch, Vanessa and Lehmann, Beat and Morari, Manfred},
  journal={Energy and buildings},
  volume={45},
  pages={15--27},
  year={2012},
  publisher={Elsevier}
}

@article{drgovna2020all,
  title={All you need to know about model predictive control for buildings},
  author={Drgo{\v{n}}a, J{\'a}n and Arroyo, Javier and Figueroa, Iago Cupeiro and Blum, David and Arendt, Krzysztof and Kim, Donghun and Oll{\'e}, Enric Perarnau and Oravec, Juraj and Wetter, Michael and Vrabie, Draguna L and others},
  journal={Annual Reviews in Control},
  volume={50},
  pages={190--232},
  year={2020},
  publisher={Elsevier}
}

@article{killian2016ten,
  title={Ten questions concerning model predictive control for energy efficient buildings},
  author={Killian, Michaela and Kozek, Martin},
  journal={Building and Environment},
  volume={105},
  pages={403--412},
  year={2016},
  publisher={Elsevier}
}

@article{stoffel2023evaluation,
  title={Evaluation of advanced control strategies for building energy systems},
  author={Stoffel, Phillip and Maier, Laura and K{\"u}mpel, Alexander and Schreiber, Thomas and M{\"u}ller, Dirk},
  journal={Energy and Buildings},
  volume={280},
  pages={112709},
  year={2023},
  publisher={Elsevier}
}

@article{pergantis2024field,
  title={Field demonstration of predictive heating control for an all-electric house in a cold climate},
  author={Pergantis, Elias N and Al Theeb, Nadah and Dhillon, Parveen and Ore, Jonathan P and Ziviani, Davide and Groll, Eckhard A and Kircher, Kevin J and others},
  journal={Applied Energy},
  volume={360},
  pages={122820},
  year={2024},
  publisher={Elsevier}
}

@article{xiao2023building,
  title={Building thermal modeling and model predictive control with physically consistent deep learning for decarbonization and energy optimization},
  author={Xiao, Tianqi and You, Fengqi},
  journal={Applied Energy},
  volume={342},
  pages={121165},
  year={2023},
  publisher={Elsevier}
}

@article{jorissen2019taco,
  title={TACO, an automated toolchain for model predictive control of building systems: implementation and verification},
  author={Jorissen, Filip and Boydens, Wim and Helsen, Lieve},
  journal={Journal of building performance simulation},
  volume={12},
  number={2},
  pages={180--192},
  year={2019},
  publisher={Taylor \& Francis}
}

@article{bunning2022physics,
  title={Physics-informed linear regression is competitive with two Machine Learning methods in residential building MPC},
  author={B{\"u}nning, Felix and Huber, Benjamin and Schalbetter, Adrian and Aboudonia, Ahmed and de Badyn, Mathias Hudoba and Heer, Philipp and Smith, Roy S and Lygeros, John},
  journal={Applied Energy},
  volume={310},
  pages={118491},
  year={2022},
  publisher={Elsevier}
}

@article{balali2023energy,
  title={Energy modelling and control of building heating and cooling systems with data-driven and hybrid models—A review},
  author={Balali, Yasaman and Chong, Adrian and Busch, Andrew and O’Keefe, Steven},
  journal={Renewable and Sustainable Energy Reviews},
  volume={183},
  pages={113496},
  year={2023},
  publisher={Elsevier}
}

@article{hu2023multi,
  title={Multi-zone building control with thermal comfort constraints under disjunctive uncertainty using data-driven robust model predictive control},
  author={Hu, Guoqing and You, Fengqi},
  journal={Advances in Applied Energy},
  volume={9},
  pages={100124},
  year={2023},
  publisher={Elsevier}
}

@article{korda2014stochastic,
  title={Stochastic MPC framework for controlling the average constraint violation},
  author={Korda, Milan and Gondhalekar, Ravi and Oldewurtel, Frauke and Jones, Colin N},
  journal={IEEE Transactions on Automatic Control},
  volume={59},
  number={7},
  pages={1706--1721},
  year={2014},
  publisher={IEEE}
}

@article{blum2021building,
  title={Building optimization testing framework (BOPTEST) for simulation-based benchmarking of control strategies in buildings},
  author={Blum, David and Arroyo, Javier and Huang, Sen and Drgo{\v{n}}a, J{\'a}n and Jorissen, Filip and Walnum, Harald Taxt and Chen, Yan and Benne, Kyle and Vrabie, Draguna and Wetter, Michael and others},
  journal={Journal of Building Performance Simulation},
  volume={14},
  number={5},
  pages={586--610},
  year={2021},
  publisher={Taylor \& Francis}
}

@article{shi2024dad,
  title={Disturbance-adaptive Model Predictive Control for Bounded Average Constraint Violations},
  author={Shi, Jicheng and Jones, Colin N},
  journal={arXiv preprint arXiv:2503.24169},
  year={2025}
}

@article{comite2007indoor,
  title={Indoor environmental input parameters for design and assessment of energy performance of buildings addressing indoor air quality, thermal environment, lighting and acoustics},
  author={Comite'Europe'en de Normalisation, CEN},
  journal={EN 15251},
  year={2007},
  publisher={CEN}
}

@article{sourbron2011evaluation,
  title={Evaluation of adaptive thermal comfort models in moderate climates and their impact on energy use in office buildings},
  author={Sourbron, Maarten and Helsen, Lieve},
  journal={Energy and Buildings},
  volume={43},
  number={2-3},
  pages={423--432},
  year={2011},
  publisher={Elsevier}
}

@article{shi2025adaptive,
  title={Adaptive data-driven prediction in a building control hierarchy: A case study of demand response in Switzerland},
  author={Shi, Jicheng and Lian, Yingzhao and Salzmann, Christophe and Jones, Colin N},
  journal={Energy and Buildings},
  pages={115498},
  year={2025},
  publisher={Elsevier}
}

@misc{united2023,
  title        = {2023 Global Status Report for Buildings and Construction},
  author       = {United Nations Environment Programme},
  year         = {2024},
  address      = {Nairobi, Kenya},
  publisher    = {United Nations Environment Programme (UNEP) and Global Alliance for Buildings and Construction (GlobalABC)},
  doi          = {10.59117/20.500.11822/45095},
}

@inproceedings{korda2012stochastic,
  title={Stochastic model predictive control: Controlling the average number of constraint violations},
  author={Korda, Milan and Gondhalekar, Ravi and Oldewurtel, Frauke and Jones, Colin N},
  booktitle={2012 IEEE 51st IEEE Conference on Decision and Control (CDC)},
  pages={4529--4536},
  year={2012},
  organization={IEEE}
}

@article{chinde2022data,
  title={Data-enabled predictive control for building HVAC systems},
  author={Chinde, Venkatesh and Lin, Yashen and Ellis, Matthew J},
  journal={J of Dyn Syst, Meas, and Control},
  volume={144},
  number={8},
  pages={081001},
  year={2022},
  publisher={American Society of Mechanical Engineers}
}

@article{yin2024data,
  title={Data-driven predictive control for demand side management: Theoretical and experimental results},
  author={Yin, Mingzhou and Cai, Hanmin and Gattiglio, Andrea and Khayatian, Fazel and Smith, Roy S and Heer, Philipp},
  journal={Appl Energy},
  volume={353},
  pages={122101},
  year={2024},
  publisher={Elsevier}
}

@article{dorfler2022bridging,
  title={Bridging direct \& indirect data-driven control formulations via regularizations and relaxations},
  author={Dorfler, Florian and Coulson, Jeremy and Markovsky, Ivan},
  journal={IEEE Trans on Autom Control},
  year={2022},

  publisher={IEEE}
}

@inproceedings{berberich2020robust,
  title={Robust constraint satisfaction in data-driven MPC},
  author={Berberich, Julian and K{\"o}hler, Johannes and M{\"u}ller, Matthias A and Allg{\"o}wer, Frank},
  booktitle={2020 59th IEEE Conference on Decision and Control (CDC)},
  pages={1260--1267},
  year={2020},
  organization={IEEE}
}

@book{cp_vovk2005,
  title={Algorithmic learning in a random world},
  author={Vovk, Vladimir and Gammerman, Alexander and Shafer, Glenn},
  volume={29},
  year={2005},
  publisher={Springer}
}

@article{cp_intro_angelopoulos2021,
  title={A gentle introduction to conformal prediction and distribution-free uncertainty quantification},
  author={Angelopoulos, Anastasios N and Bates, Stephen},
  journal={arXiv preprint arXiv:2107.07511},
  year={2021}
}

@inproceedings{fleming2017time,
  title={Time-average constraints in stochastic Model Predictive Control},
  author={Fleming, James and Cannon, Mark},
  booktitle={2017 American Control Conference (ACC)},
  pages={5648--5653},
  year={2017},
  organization={IEEE}
}

@inproceedings{oldewurtel2013adaptively,
  title={Adaptively constrained stochastic model predictive control for closed-loop constraint satisfaction},
  author={Oldewurtel, Frauke and Sturzenegger, David and Esfahani, Peyman Mohajerin and Andersson, G{\"o}ran and Morari, Manfred and Lygeros, John},
  booktitle={2013 American Control Conference},
  pages={4674--4681},
  year={2013},
  organization={IEEE}
}

@incollection{bemporad2007robust,
  title={Robust model predictive control: A survey},
  author={Bemporad, Alberto and Morari, Manfred},
  booktitle={Robustness in identification and control},
  pages={207--226},
  year={2007},
  publisher={Springer}
}

@article{mesbah2016stochastic,
  title={Stochastic model predictive control: An overview and perspectives for future research},
  author={Mesbah, Ali},
  journal={IEEE Control Systems Magazine},
  volume={36},
  number={6},
  pages={30--44},
  year={2016},
  publisher={IEEE}
}

@article{lian2023adaptive,
  title={Adaptive robust data-driven building control via bilevel reformulation: An experimental result},
  author={Lian, Yingzhao and Shi, Jicheng and Koch, Manuel and Jones, Colin Neil},
  journal={IEEE Transactions on Control Systems Technology},
  volume={31},
  number={6},
  pages={2420--2436},
  year={2023},
  publisher={IEEE}
}

@inproceedings{shi2023efficient,
  title={Efficient Recursive Data-enabled Predictive Control},
  author={Shi, Jicheng and Lian, Yingzhao and Jones, Colin N},
  booktitle={2024 European Control Conference (ECC)},
  year={2024},
  organization={IEEE}
}

@inproceedings{di2022lessons,
  title={Lessons learned from data-driven building control experiments: Contrasting gaussian process-based mpc, bilevel deepc, and deep reinforcement learning},
  author={Di Natale, Loris and Lian, Yingzhao and Maddalena, Emilio T and Shi, Jicheng and Jones, Colin N},
  booktitle={2022 IEEE 61st conference on decision and control (CDC)},
  pages={1111--1117},
  year={2022},
  organization={IEEE}
}

@article{willems2005note,
  title={A note on persistency of excitation},
  author={Willems, Jan C and Rapisarda, Paolo and Markovsky, Ivan and De Moor, Bart LM},
  journal={Systems \& Control Letters},
  volume={54},
  number={4},
  pages={325--329},
  year={2005},
  publisher={Elsevier}
}

@misc{deru2008doe,
    key = {BOPTEST},
  title        = {BOPTEST Test Case: Multizone Office Simple Air},
  howpublished = {\url{https://ibpsa.github.io/project1-boptest/docs-testcases/multizone_office_simple_air/index.html}},
  note         = {Accessed: 2026-02-18}
}

@article{o2022data,
  title={Data-driven predictive control with improved performance using segmented trajectories},
  author={O’Dwyer, Edward and Kerrigan, Eric C and Falugi, Paola and Zagorowska, Marta and Shah, Nilay},
  journal={IEEE Transactions on Control Systems Technology},
  volume={31},
  number={3},
  pages={1355--1365},
  year={2022},
  publisher={IEEE}
}

@article{
swales1999open,
  title={Open $\text{M}$odbus/$\text{TCP}$ Specification},
  author={Swales, Andy and others},
  journal={Schneider Electric},
  volume={29},
  pages={3--19},
  year={1999}
}

@misc{influxdb,
key = {influxdb},
  title = {InfluxDB},
  howpublished = {\url{https://www.influxdata.com}},
  note = {Accessed: 2025-05-01}
}

@misc{tomorrow,
key = {tomorrow},
title = {Tomorrow.io},
howpublished={\url{https://www.tomorrow.io}},
note = {Accessed: 2025-05-01},
}

@misc{emu,
key = {emu},
title = {EMU Electronics AG},
howpublished={\url{https://www.emuag.ch/}},
note = {Accessed: 2025-05-01},
}

@misc{boptestChallenge,
key = {boptest},
title = {Adrenalin BOPTEST Challenge:
Smart building HVAC control},
howpublished={\url{https://adrenalin.energy/BOPTEST-Challenge-Smart-building-HVAC-control}},
note = {Accessed: 2025-11-05},
}

@misc{matlab:quadprog,
  title        = {MATLAB Optimization Toolbox: quadprog function},
  howpublished = {\url{https://www.mathworks.com/help/optim/ug/quadprog.html}},
  note         = {Accessed: 2025-11-05}
}

@article{ghosh2025adaptive,
  title={Adaptive Relaxation-Based Nonconservative Chance Constrained Stochastic MPC},
  author={Ghosh, Avik and Cortes-Aguirre, Cristian and Chen, Yi-An and Khurram, Adil and Kleissl, Jan},
  journal={IEEE Transactions on Control Systems Technology},
  year={2025},
  publisher={IEEE}
}

@misc{gurobi,  author = {{Gurobi Optimization, LLC}},  title = {{Gurobi Optimizer Reference Manual}},  year = 2024,  url = "https://www.gurobi.com"}

@article{alanwar2022robust,
  title={Robust data-driven predictive control using reachability analysis},
  author={Alanwar, Amr and St{\"u}rz, Yvonne and Johansson, Karl Henrik},
  journal={European Journal of Control},
  volume={68},
  pages={100666},
  year={2022},
  publisher={Elsevier}
}

@inproceedings{de2024koopman,
  title={Koopman data-driven predictive control with robust stability and recursive feasibility guarantees},
  author={de Jong, Thomas and Breschi, Valentina and Schoukens, Maarten and Lazar, Mircea},
  booktitle={2024 IEEE 63rd Conference on Decision and Control (CDC)},
  pages={140--145},
  year={2024},
  organization={IEEE}
}

@article{dubied2025robust,
  title={A robust and adaptive MPC formulation for Gaussian process models},
  author={Dubied, Mathieu and Lahr, Amon and Zeilinger, Melanie N and K{\"o}hler, Johannes},
  journal={arXiv preprint arXiv:2507.02098},
  year={2025}
}

@article{schimperna2024robust,
  title={Robust constrained nonlinear model predictive control with gated recurrent unit model},
  author={Schimperna, Irene and Magni, Lalo},
  journal={Automatica},
  volume={161},
  pages={111472},
  year={2024},
  publisher={Elsevier}
}

@article{gao2023energy,
  title={Energy saving and indoor temperature control for an office building using tube-based robust model predictive control},
  author={Gao, Yuan and Miyata, Shohei and Akashi, Yasunori},
  journal={Applied Energy},
  volume={341},
  pages={121106},
  year={2023},
  publisher={Elsevier}
}

@article{maddalena2022experimental,
  title={Experimental data-driven model predictive control of a hospital HVAC system during regular use},
  author={Maddalena, Emilio T and Mueller, Silvio A and dos Santos, Rafael M and Salzmann, Christophe and Jones, Colin N},
  journal={Energy and Buildings},
  volume={271},
  pages={112316},
  year={2022},
  publisher={Elsevier}
}

@article{li2022tube,
  title={Tube-based robust model predictive control of multi-zone demand-controlled ventilation systems for energy saving and indoor air quality},
  author={Li, Bingxu and Wu, Bingjie and Peng, Yelun and Cai, Wenjian},
  journal={Applied Energy},
  volume={307},
  pages={118297},
  year={2022},
  publisher={Elsevier}
}

@article{mahmood2023data,
  title={Data-driven robust model predictive control for greenhouse temperature control and energy utilisation assessment},
  author={Mahmood, Farhat and Govindan, Rajesh and Bermak, Amine and Yang, David and Al-Ansari, Tareq},
  journal={Applied Energy},
  volume={343},
  pages={121190},
  year={2023},
  publisher={Elsevier}
}

\end{document}